\documentclass[12pt,preprint]{aastex}

\slugcomment{Accepted ApJ Submission: 8/2/04}

\shorttitle{KISS Metal Abundances III.}
\shortauthors{Lee, Salzer \& Melbourne}

\begin{document}


\title{Metal Abundances of KISS Galaxies\\ III. Nebular Abundances for Fourteen Galaxies and \\ the Luminosity-Metallicity Relationship for \ion{H}{2} Galaxies}


\author{Janice C. Lee}
\affil{Steward Observatory, University of Arizona, Tucson, AZ 85712}
\email{jlee@as.arizona.edu}

\author{John J. Salzer}
\affil{Astronomy Department, Wesleyan University, Middletown, CT 06459}
\email{slaz@astro.wesleyan.edu}

\and

\author{Jason Melbourne}
\affil{UCO/Lick Observatory, UC Santa Cruz, Santa Cruz, CA 95064}
\email{jmel@ucolick.org}


\begin{abstract}

We report results from the third in a series of nebular abundance studies of emission-line galaxies from the KPNO International Spectroscopic Survey (KISS).  Galaxies with coarse metallicity estimates of 12 + log(O/H) less than 8.2 dex were selected for observation.  Spectra of 14 galaxies, which cover the full optical region from [OII]$\lambda\lambda$3727,3729 to beyond [SII]$\lambda\lambda$6717,6731, are presented, and abundance ratios of N, O, Ne, S, and Ar are computed.  The auroral [OIII]$\lambda$4363 line is detected in all 14 galaxies.  Oxygen abundances determined through the direct electron temperature ($T_e$) method confirm that the sample is metal-poor with 7.61 $\leq$ 12 + log(O/H) $\leq$ 8.32.  By using these abundances in conjunction with other $T_e$-based measurements from the literature, we demonstrate that \ion{H}{2} galaxies and more quiescent dwarf irregular galaxies follow similar metallicity-luminosity (L-Z) relationships.  The primary difference is a zero-point shift between the correlations such that \ion{H}{2} galaxies are brighter by an average of 0.8 B magnitudes at a given metallicity.  This offset can be used as evidence to argue that low-luminosity \ion{H}{2} galaxies typically undergo factor of two luminosity enhancements, and starbursts that elevate the luminosities of their host galaxies by 2 to 3 magnitudes are not as common.  We also demonstrate that the inclusion of interacting galaxies can increase the scatter in the L-Z relation and may force the observed correlation towards lower metallicities and/or larger luminosities.  This must be taken into account when attempting to infer metal abundance evolution by comparing local L-Z relations with ones based on higher redshift samples since the fraction of interacting galaxies should increase with look-back time.

\end{abstract}


\section{Introduction}
One of the broad goals of the KPNO International Spectroscopic Survey (KISS, Salzer et al. 2000) is to provide a large, statistically complete sample of emission-line galaxies (ELGs) that can be used to study the chemical enrichment and star-formation properties of active galaxies in the local universe (z $\lesssim$ 0.095).  As part of this endeavor, we have been engaged in a long-term effort to obtain optical long-slit spectroscopy for KISS ELG candidates identified via line-emission in the initial low-dispersion objective-prism survey.  The emission lines in the spectra from these follow-up observations supply accurate redshifts, allow for the determination of the type of activity occurring in the galaxies (Baldwin, Phillips \& Terlevich 1981, Veilleux \& Osterbrock 1987) and depending on the quality of the data, can also provide a measurement of the heavy element abundance (e.g. Searle \& Sargent 1972, Izotov, Thuan \& Lipovetsky 1994) and the star formation rate (Kennicutt 1998).  

Metal abundance, in particular, provides insight into the evolutionary status of galaxies because the presence of elements heavier than hydrogen and helium indicates that chemical processing has occurred via star formation and the subsequent expulsion of metals into the interstellar medium.  By combining abundance measurements with additional quantities such as luminosity and gas mass, a wealth of analyses may be undertaken.  For example, models of chemical evolution which can constrain the presence of inflows and outflows may be tested (e.g. Kennicutt \& Skillman 2001).  Correlations of the metallicity with other properties such as dynamical mass (e.g. Lequeuex et al. 1979), rotational velocity (e.g. Garnett 2002), and B luminosity (e.g. Skillman, Kennicutt \& Hodge 1989) may be studied.  The application of the last of these analyses to large datasets has recently begun (e.g. 2dF, Lamareille et al. 2004; SDSS, Tremonti et al. 2004), and includes the work presented in this series of papers describing our follow-up abundance studies of KISS star-forming ELGs.  In these papers, we have thus far: (1) empirically derived coarse abundance estimates for 519 galaxies and used these results to determine the form of the B-band metallicity-luminosity relationship over approximately seven magnitudes (Melbourne \& Salzer 2002; hereafter Paper I), and (2) begun to publish the follow-up spectra and present nebular abundance calculations for high signal-to-noise data which include detection of the temperature sensitive [\ion{O}{3}]$\lambda$4363 line (Melbourne et al. 2004; hereafter Paper II). 

In this paper, we present abundance quality spectrophotometry for an additional 14 galaxies collected over two seasons at the 6.5m MMT\footnote{The MMT Observatory is a joint facility of the University of Arizona and the Smithsonian Institution.} on Mount Hopkins.  The galaxies chosen for these observations are ones which are identified as metal-poor (12 + log(O/H) $\lesssim$ 8.2) using the empirical strong-lined estimator developed in Paper I.  Our MMT targets previously had ``quick-look'' follow-up spectra available (most of which were taken with spectrographs with little or no sensitivity in the blue), which allowed for the calculation of these coarse estimates based on the [\ion{O}{3}]$\lambda$5007/H$\beta$ and [\ion{N}{2}]$\lambda$6583/H$\alpha$ line ratios, but not for the more accurate $T_e$-based nebular abundances which require the [\ion{O}{3}]$\lambda$4363 line.  Therefore, one of the objectives of the MMT observations was to secure robust detections of [\ion{O}{3}]$\lambda$4363 for additional targets in order to enlarge the sample of 12 galaxies presented in Paper II.  Combining the data presented here and in Paper II, there are 23 unique KISS ELGs for which $T_e$-based metallicities can be computed.  These results are used in the next paper in the KISS abundance series (Salzer et al. 2004, hereafter Paper IV) to improve the calibration of our empirical strong-lined oxygen abundance estimator, to update the KISS ELG L-Z relation in the B-band with a larger sample of 763 galaxies, and to further determine it in the near-infrared J, H and K-bands. 

In the analysis given here, our sample of 23 KISS galaxies with directly determined nebular abundances is used in conjunction with other data from the literature to re-evaluate the B-band luminosity-metallicity relationship for low-luminosity \ion{H}{2} galaxies, and to compare this relation with the one followed by the population of more quiescent dwarf irregular (dIrr) galaxies.  While a tight correlation of increasing oxygen abundance with increasing luminosity has already been carefully quantified by exclusively using $T_e$-determined abundances and accurate photometry for the dIrrs (e.g. Skillman et al. 1989, Richer \& McCall 1995, H. Lee et al. 2003), a similar treatment for \ion{H}{2} galaxies is not available.  In fact, previous investigations of the L-Z relation in the low-luminosity regime have typically excluded galaxies that are strongly star-forming since these objects are thought to obscure any relationship that might be found.  An implicit assumption here is that the underlying relationship is between mass and metallicity (e.g. Lequeux et al. 1979, Tremonti et al. 2004), and that the large contribution of light from starbursting regions causes the observed B luminosity to be a poor indicator of stellar mass.  Papers which do investigate the L-Z relationship for \ion{H}{2} galaxies have primarily examined it using either small samples, empirically estimated abundances, and/or estimated luminosities (e.g. Hunter \& Hoffman 1999,  Salzer et al. 1989, Kobulnicky \& Zaritsky 1999).  In general, these studies also find a trend of increasing metallicity with increasing luminosity for the \ion{H}{2} galaxies, albeit with considerably larger scatter, and sometimes with major deviations when compared to the L-Z relationship for dIrrs.  Here, we attempt to provide a more robust analysis for this population.  {\it Using only directly determined abundances}, we find that \ion{H}{2} galaxies and more quiescent dIrrs follow similar relationships.  We quantify and discuss some possible physical interpretations of the differences that exist.

\section{Spectroscopy}

\subsection{Properties of the Observed Sample}
General properties of the galaxies that were observed at the MMT are compiled in Table~\ref{tab1}.  The objects selected for observation are ones with 12 + log(O/H) $\lesssim$ 8.2 dex, as determined via the empirical strong-lined estimator developed in Paper I.  Nevertheless, the observed sample still spans a wide range in luminosity, from -12.46 to -19.40 in $M_B$.  We also preferentially chose targets with lower apparent brightnesses to take advantage of the MMT's large light collecting area.  The median $m_B$ of the observed sample is 18.40.

\begin{figure}
\plotone{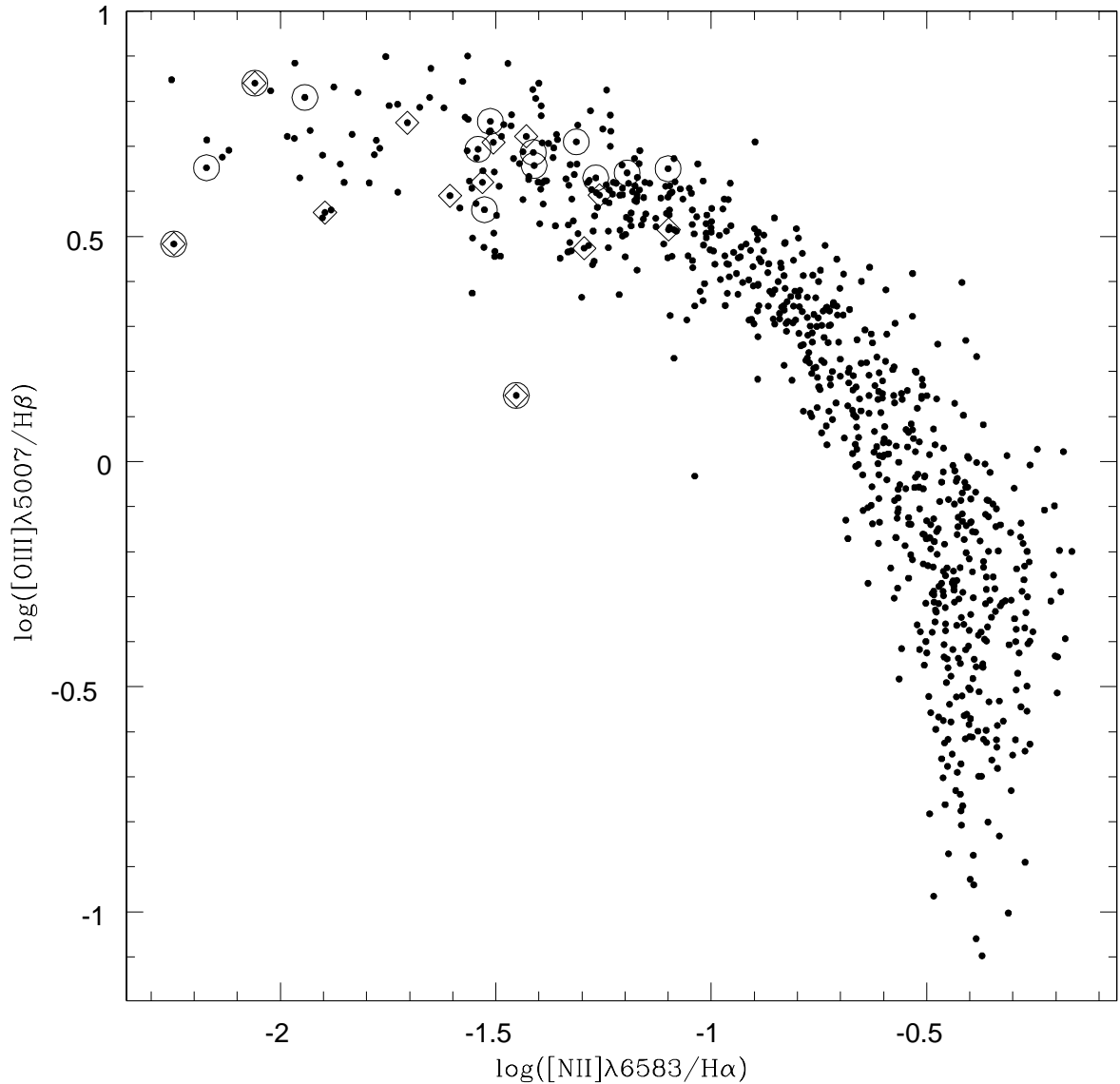}
\figcaption{Line diagnostic diagram for KISS galaxies determined to have emission powered by hot OB stars.  Star-forming ELGs that have ``quick-look'' follow-up spectra with quality code 1 or 2 in the KISS database (see text) are shown as points.  Galaxies with nebular abundance derived from spectra taken at the Lick 3m (Paper II) are marked by open diamonds while galaxies with nebular abundance derived from spectra taken at the MMT (this paper) are marked by open circles.  Three objects were re-observed at the MMT to check dubious line fluxes derived from the Lick spectra. \label{diag_diagram}}
\end{figure}

The locations of the targets in the log([\ion{O}{3}]$\lambda$5007/H$\beta$) versus log([\ion{N}{2}]$\lambda$6583/H$\alpha$) plane are shown in Figure~\ref{diag_diagram}.  This classical diagnostic diagram (Baldwin, Phillips \& Terlevich 1981) is used to empirically distinguish between ELGs in which photoionization is primarily due to non-thermal, power-law continua (i.e. ones which contain AGN), and those where the emission is mainly powered by hot OB stars.   In this figure, the points represent all KISS ELGs with good quality spectra\footnote{Spectra are assigned one of three quality codes in the KISS database.  $Q = 1$ refers to high-quality spectra, with high S/N emission lines and reliable emission-line ratios, although a high S/N detection of O[III]$\lambda$4363 is not necessarily implied. $Q = 2$ is assigned to lower S/N spectra that still have reliable line-ratios.   $Q = 3$ refers to the lowest-quality spectra, which usually belong to the faintest objects in the sample or to objects with intrinsically weak lines.  Only galaxies that have quality codes 1 and 2 are plotted in Figure~\ref{diag_diagram}.} which have been classified as star-forming.  Galaxies with nebular abundances based on spectra taken at Lick Observatory (Paper II) are additionally marked with open diamonds while the ones presented here are circled.  Three galaxies observed at Lick (KISSR 85, KISSR 666 and KISSB 23) were re-observed with the MMT in order to check dubious [\ion{O}{3}]$\lambda$4363 detections as discussed in detail in Paper II.

The two strong-line ratios plotted in Figure~\ref{diag_diagram} also form the basis of the empirical abundance determination method described in Paper I.  For star-forming systems, the distribution of points in this plot form a well-defined locus: low-metallicity systems occupy the upper left and the metal abundance smoothly increases as one moves along the locus toward the lower right corner of the diagram.  As would be expected, all of the MMT targets lie within the low-metallicity, high-excitation region of the plot; they generally have high [\ion{O}{3}]$\lambda$5007/H$\beta$ ratios but low [\ion{N}{2}]$\lambda$6583/H$\alpha$ ratios.  The one galaxy which lies considerably off the locus, and has been observed both at Lick and the MMT, is KISSB 23.  As discussed in Paper II, the emission in this object is powered by a lower excitation source, most likely because its starburst is past its peak, as indicated by a relatively low value of [OIII]$\lambda$ 5007/[OII]$\lambda$ 3727 calculated from its spectrum.

\subsection{Observations}
The data presented in this paper were obtained with the Blue Channel Spectrograph on the 6.5m MMT on the nights of May 10, 2002 and April 3, 2003.  Conditions were photometric on both nights.  In May 2002, we used a 500 line mm$^{-1}$ grating blazed at 5410\AA\ in first order along with a 3600\AA\ UV blocking filter to prevent overlap from second order light in the red.  With the 3072 $\times$ 1024 15$\mu$ pixel CCD detector (CCD 22) installed on the spectrograph at that time, this set-up provided coverage between 3650\AA\ and 7200\AA\ , a pixel scale of 0\farcs6 (when binned by 2 in the spatial direction), and a spectral resolution of 5\AA\ FWHM for a 1\farcs5 $\times$ 180\arcsec\ slit.  At the end of February 2003, the detector was replaced with a smaller 2688 $\times$ 512 15$\mu$ pixel chip (CCD 35), resulting in a loss of approximately 400\AA\ in spectral coverage when the above set-up is used.  Therefore, in April 2003 we chose to use a lower dispersion 300 line mm$^{-1}$ grating instead to ensure that wavelengths between and including [\ion{O}{2}]$\lambda\lambda$3726,29 and [\ion{S}{2}]$\lambda\lambda$6717,31 were adequately sampled by the detector for the range of redshifts spanned by our targets ($\Delta cz\sim24,000$ km s$^{-1}$).   The 300 line mm$^{-1}$ grating, blazed at 4800\AA\ and used in first order, provided coverage between 3600\AA\ and 8400\AA\ , and a spectral resolution of 9\AA\ FWHM for a 1\farcs5 $\times$ 180\arcsec\ slit.  The spatial scale did not change since the pixel sizes on the old and new detectors were the same.

Exposure times for each galaxy ranged from 10 to 15 minutes (see Table~\ref{tab1}).  Our objective was to obtain good signal-to-noise (SNR$\sim$20) in the weak [\ion{O}{3}]$\lambda$4363 line so that it could be used to estimate the nebular electron temperature which is necessary for an accurate determination of the metal abundance.  Since the sensitivity of the slit-viewing camera allowed for direct viewing of the majority our targets, we were able to confirm that the pointing of the telescope was accurate enough to allow for slit positioning based on the target's coordinates (as listed in Table~\ref{tab1}), and were also able to make small adjustments to the pointing by eye when necessary. 

Spectrophotometric standards from Massey et al. (1988) were observed every few hours to allow for flux calibration.  The spectrograph slit was aligned with the parallactic angle to minimize the effects of differential atmospheric refraction for all standard star and galaxy observations.  Bias, HeNeAr lamp, quartz lamp and twilight flat-field exposures were taken following routine procedures.    

\subsection{Data Reduction}
The data were reduced following standard methods, using Image 
Reduction and Analysis Facility\footnote{IRAF is distributed by
the National Optical Astronomy Observatories, which are operated 
by AURA, Inc. under cooperative agreement with the National 
Science Foundation.} (IRAF) software.  The bias level was determined
for each frame using an overscan region, while a mean bias image
constructed from 20 zero-second exposures was used to correct for
any two-dimensional structure introduced by the read-out electronics.
The dark-count level was measured using a series of exposures taken
with the shutter closed, and was found to be negligible.  Flat-fielding
was carried out by combining a series of several quartz-lamp spectra
that was normalized and corrected for any wavelength-dependent response.
Spectra of the twilight sky were used to account for the illumination 
variation along the slit.  These variations were found to be very small (under 1\%).

Once the standard CCD reductions were completed, the spectra were extracted
to a one-dimensional format using the APALL routine.  The extraction
apertures were set on a case-by-case basis, and depended on the two-dimensional
profile of the source along the slit.  In most cases, the emission regions
of our targets were unresolved spatially, and the extraction region was 
chosen to include most of the line flux present (8-11 pixels or 4\farcs8-6\farcs6 wide).  The sky background was subtracted in the same step, typically using
object-free regions 20-30 pixels wide on either side of the target spectrum.
The He-Ne-Ar lamp spectra were used to establish the wavelength scale, and
spectra of several spectrophotometric standard stars from Massey et al. (1988) were used to 
establish the flux scale.  The standard stars were also used to create a
template telluric absorption spectrum that was scaled and used to correct the 
spectra of our ELGs.  This step is important, since at the typical redshifts 
of our target galaxies lines like [\ion{S}{2}]$\lambda\lambda$6717, 6731,
[\ion{N}{2}]$\lambda\lambda$6548, 6583, and/or H$\alpha$ can fall within the
atmospheric B-band (6860-6890\AA) and have their fluxes significantly 
underestimated.  In cases where
more than one spectral image was obtained for a given galaxy, the individual
spectra were processed fully and then combined prior to measurement.  All
line fluxes were measured using the SPLOT routine.

        An accurate estimate of the internal reddening along the line-of-sight to each emission region is measured using the Balmer line ratios.  We use a routine that simultaneously solves for the underlying Balmer absorption and for the exponential reddening coefficient $c_{H\beta}$.  The three lowest-order Balmer line ratios are used for this process.  The values for $c_{H\beta}$ and the equivalent width of the underlying Balmer absorption lines (assumed to be the same for all four lines) that give consistent results for all three ratios are determined.  For the majority of the galaxies observed we find an underlying absorption equivalent width of 3--4 \AA.  The derived estimates of $c_{H\beta}$ for each galaxy are shown in Table 2.  This value of $c_{H\beta}$ is then used to correct all measured line ratios for reddening, following the standard prescription (e.g., Osterbrock(1989):
\begin{equation}
\frac{I(\lambda)}{I(\mbox{H}\beta)}=
\frac{F(\lambda)}{F(\mbox{H}\beta)}exp[c_{H\beta}f(\lambda)]
\end{equation}
where f($\lambda$) is derived from studies of absorption in the Milky Way (using values taken from Rayo et al. 1982).

\subsection{Spectral Data}
\begin{figure}
\epsscale{1.5}
\vspace*{-0.5in}
\hspace*{-0.7in}
\plotone{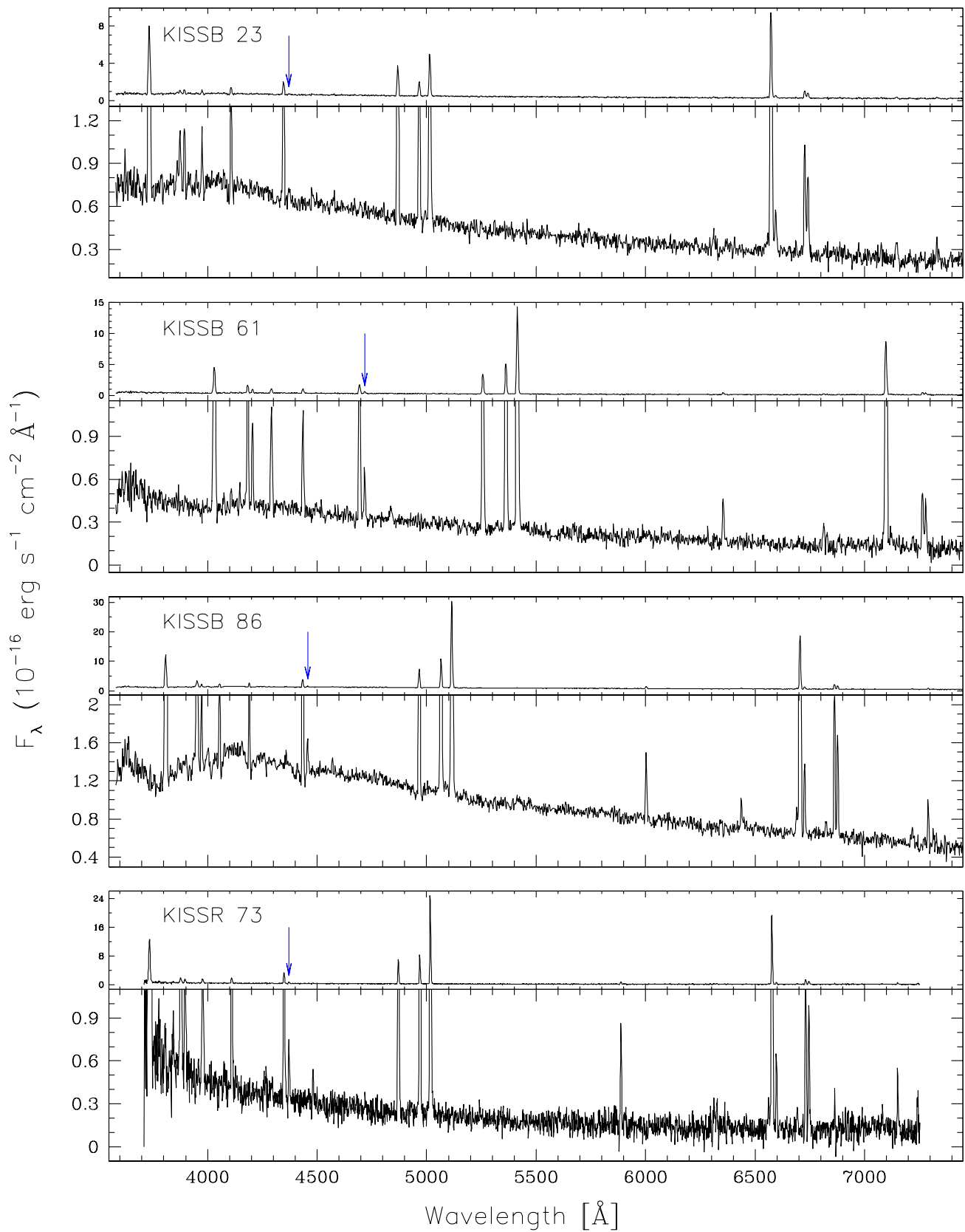}
\vspace*{-0.9in}
\figcaption{Spectra of KISS galaxies observed at the MMT.  For each galaxy, the full intensity range is plotted in the top panel to illustrate the ratios of the strong emission lines, while an expanded version is given in the bottom panel to more clearly show the quality of the weakest lines.  The location of the [OIII]$\lambda$4363 line is marked by an arrow in the upper panel.\label{spectra}}
\end{figure}

Spectra of the 14 galaxies observed at the MMT are presented in Figure~\ref{spectra}.  As would be expected for the high-equivalent width, star-forming ELGs found in KISS, the galaxies in this sample exhibit \ion{H}{2} region-like spectra which have strong emission-lines superimposed on a faint, blue continuum.  The full intensity range is plotted for each galaxy in the top panel to illustrate the ratios of the strong emission-lines, while an expanded version is presented in the bottom panel to more clearly show the quality of the weakest lines.  Reddening corrected flux ratios with respect to H$\beta$ for lines with equivalent widths above 1\AA\ are reported in Table~\ref{tab2}.

\section{Metal Abundances}
We calculate metal abundances from the MMT spectra, following the same techniques described in Paper II.  The method is summarized below and our results are presented in Table~\ref{tab3}.

\subsection{Electron Density and Temperatures}
Calculations of the electron density and electron temperature are performed by using the IRAF NEBULAR package ZONES routine (Shaw et al. 1995).  We assume a two zone ionization model for the nebular emission regions studied.  In the higher ionization zone oxygen is doubly ionized, whereas in the low ionization zone oxygen is singly ionized.  Hydrogen is assumed to be ionized throughout both regions.

We use the sulfur line ratio, [\ion{S}{2}]$\lambda$6716/$\lambda$6731 (Izotov et al. 1994) to determine the electron density, which in all measurable cases is roughly 100 e$^-$cm$^{-3}$.  In some cases the observed line ratio results in an unphysically small value of the electron density.  This is usually caused by the sulfur lines falling within the atmospheric B-band.  In these cases, we assign the galaxy the typical nebular density of 100 e$^-$cm$^{-3}$.  The sulfur lines are the only adequate electron density indicator available in the spectra, therefore we assume that both zones have the density indicated by the sulfur lines.

We calculate the temperature in the higher ionization zone using the traditional oxygen line ratio [\ion{O}{3}] ($\lambda4959 + \lambda5007$)/$\lambda$4363 (Aller 1984, Osterbrock 1989).  We then estimate the temperature in the low ionization zone, using the relationship derived by Pagel et al. (1992) based on the models of Stasinska (1990),

\begin{equation}
\label{O2temp}
t_e([OII]) = 2((t_e([OIII])^{-1} + 0.8)^{-1},
\end{equation}
where t's are temperatures measured in units of $10^4$ K.  The measured electron densities and temperatures are given in Table~\ref{tab3}.

\subsection{Ionic and Total Abundances}
The ionic abundances are calculated with the IRAF NEBULAR package ABUND (Shaw et al. 1995) routine.  The input data include the electron density and temperature for the two ionization zones, as well as the de-reddened line ratios with respect to H$\beta$.  When there is adequate S/N in the required emission-lines, we calculate ionic abundance ratios with respect to H$^+$ for the following ions: O$^+$,O$^{++}$, N$^+$, S$^+$, S$^{++}$, Ne$^{++}$, and Ar$^{++}$. 

We use the ionization correction factors given by Izotov et al. (1994),
\begin{eqnarray}
ICF(N) &=&\frac{N}{N^+}=\frac{O}{O^+},\\
ICF(Ne) &=& \frac{Ne}{Ne^{++}}=\frac{O}{O^{++}},\\
ICF(S) & = & \frac{S}{S^+ + S^{++}} \nonumber\\
       & = & [0.013 + x\{5.10 + x[-12.78 + x(14.77 - 6.11x)]\}]^{-1},\\
ICF(Ar)& = & \frac{Ar}{Ar^{++}} \nonumber\\
       & = & [0.15 + x(2.39 - 2.64x)]^{-1},\\
x & = & \frac{O^+}{O}.
\end{eqnarray}

We present total and ionic abundances for O, N, Ne, S and Ar in Table~\ref{tab3}.  For all 14 galaxies we calculate the total abundance ratio O/H.  Measurements of S/H are limited to the 6 galaxies with spectra which have S/N~$>$~5 detections of [SIII]$\lambda$6312.  We compute Ne/H, Ar/H and N/H only for those galaxies with S/N~$>$~5 in the relevant lines.  The numbers of galaxies for which we are able to calculate these quantities are given in Table~\ref{avg_ratio}.

Error estimates of the abundances are calculated by running the IRAF NEBULAR routines a second time.  We first provide the ZONES routine with oxygen line ratios which are offset by their 1 $\sigma$ error bars to generate new nebular temperatures.  We then run the ABUND routine with the original line ratios and densities and the new electron temperatures to generate ``offset'' abundances.  We take the difference between the original abundances and these ``offset'' abundances as the error in the abundance.  We add in quadrature the relative error in a given line ratio to the corresponding abundance error.  This procedure accounts for the additional error associated with low signal-to-noise ratio lines such as [\ion{N}{2}]$\lambda$6583.

\subsection{Basic Abundance Results}
Oxygen abundances determined through the direct $T_e$ method confirm that the sample of 14 galaxies observed at the MMT is relatively metal poor with 7.61 ($Z_{\odot}/20$) $\leq$ 12 + log(O/H) $\leq$ 8.32 ($Z_{\odot}/4$) \footnote{When metallicities relative to the solar value are discussed, the older value of the solar oxygen abundance of 8.92 (Lambert 1978) is assumed rather than the more recent determination of 8.69 (Allende-Prieto, Lambert \& Apslund 2001) to facilitate comparison with previous work.}.  The mean metallicity of the observed sample is 7.99 ($Z_{\odot}/8$).  The errors in the oxygen abundances range from 0.03 to 0.09 dex, with an average value of 0.04 dex.  Comparison of these $T_e$-based abundances with those computed using our strong-line estimator are presented in Paper IV. 

We now turn to examining the behavior of metal abundances relative to oxygen.  Abundance ratios have been well studied both theoretically and observationally since they provide important insights into stellar nucleosynthesis and chemical evolution in general (see review given by Pagel 1997 and references therein).  

\begin{figure}
\epsscale{1.0}
\plotone{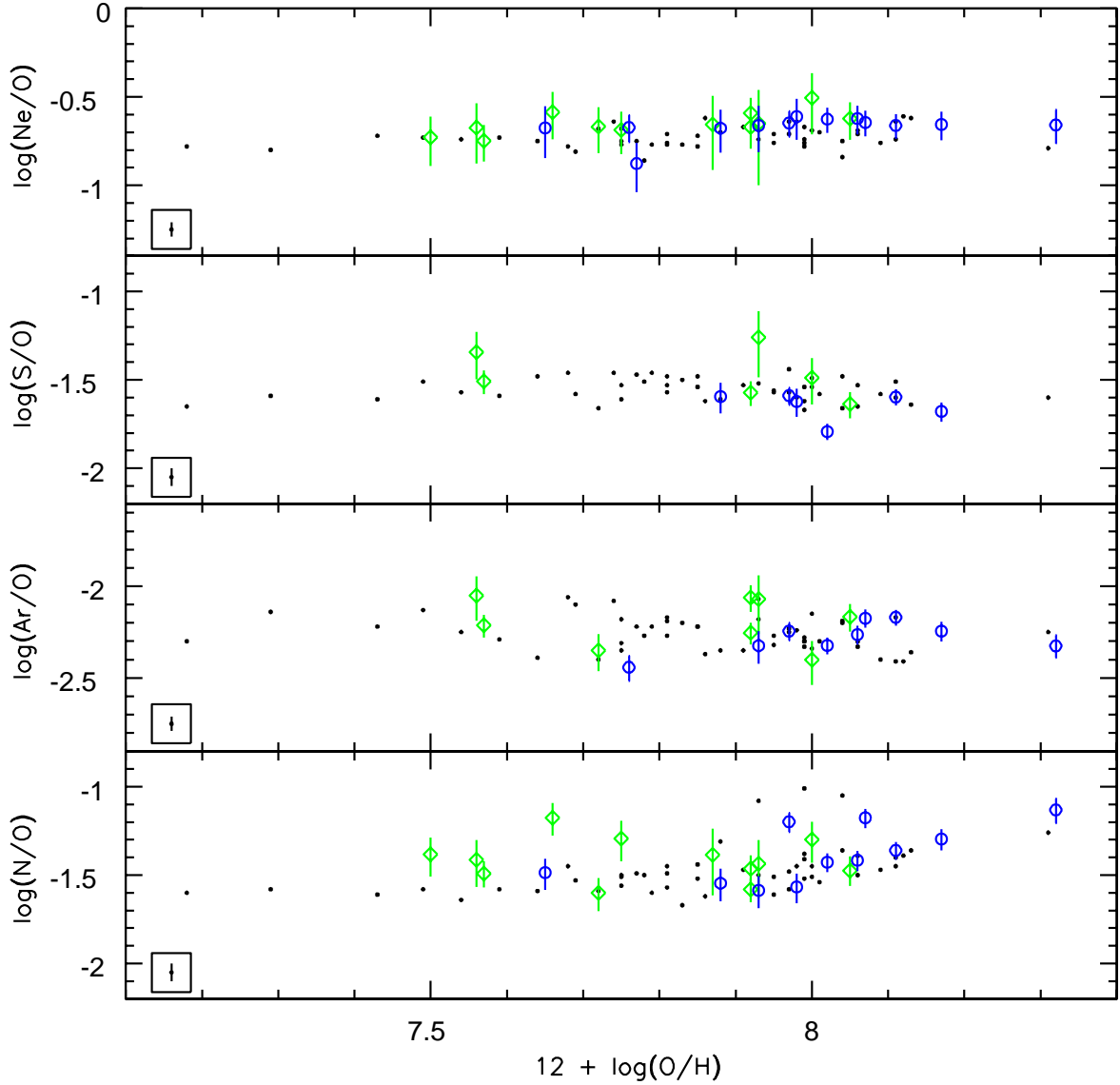}
\figcaption{Abundance ratios plotted against oxygen abundance for KISS ELGs observed at the MMT (circles, this paper), KISS ELGs observed at Lick (diamonds, Paper II) and the Izotov and Thuan (1999) sample of blue compact galaxies (filled squares).  The insets in the lower left corner of each panel shows the average errors for the Izotov and Thuan (1999) data. \label{abund_ratio}}
\end{figure}

Specifically, it is expected that the $\alpha$-process elements, such as O, Ne, S, and Ar should not vary with respect to one another since they are all primary elements created in the same population of massive stars, and should be produced in fixed proportions over a given population and promptly returned to the ISM through Type II supernovae.  The prediction of constant $\alpha$-element ratios has been borne out by previous observations (e.g. Izotov \& Thuan 1999, Paper II), and is re-confirmed by the new abundance measurements presented here.  In Figure~\ref{abund_ratio}, we plot log(Ne/O), log(S/O) and log(Ar/O) as a function of 12 + log(O/H), where the KISS ELGs observed at the MMT are indicated by open circles and those observed at Lick (Paper II) are indicated by open diamonds.  We also plot the sample of blue compact dwarfs (BCDs) presented in Izotov \& Thuan (1999) for comparison (small filled squares).  The insets in the lower left corners of each panel shows the average errors for the Izotov and Thuan (1999) data.  Our average errors are larger than those for the Izotov and Thuan (1999) dataset because of the relative faintness of our targets and the lower S/N of our spectral data.  The KISS data show no significant trends of these quantities with the oxygen abundance.   

Unlike Ne/O, S/O and Ar/O, N/O has been shown to systematically increase with oxygen abundance for higher metallicity galaxies (e.g. Vila-Costas \& Edmunds 1993).  The interpretation of this trend and its scatter in the context of understanding the dominant population of stars responsible for the build-up of N in the ISM is not straightforward and continues to be debated.  The complexity which makes the origins of N enhancement more difficult to understand is the additional mechanism of secondary N production via the CNO cycle which is expected to mainly occur in intermediate mass stars.  The reader is referred to Henry et al. (2000) for a review of the theories of N enhancement.  We simply note here that constant values of N/O favor models in which N is created as result of primary CNO processing, whereas increasing values of N/O with oxygen abundance instead indicate a contribution from secondary processing where C and O from a previous generation of stars have been incorporated.  

N/O as a function of 12 + log(O/H) is also shown in Figure~\ref{abund_ratio}.  Because the scatter at a given O/H is large for the KISS galaxies, it is difficult to discern any significant trends.  However, at 12 + log(O/H) $\gtrsim$ 8.0 the lower envelope defined by the KISS ELGs does appear to be rising linearly and follows the same relationship as the Izotov \& Thuan (1999) sample.  This may indicate secondary N production in the higher metallicity KISS ELGs.  The absence of a correlation at lower metallicities is consistent with the conclusion that N is the mainly the product of primary processes in these galaxies.  

In Table~\ref{avg_ratio}, average abundance ratios for the KISS ELGs observed at the MMT are compared with the values computed for the Lick dataset (Paper II) as well as those given in Izotov \& Thuan (1999) for their sample of BCDs.  Total averages for the combined Lick/MMT sample, where the MMT abundances supersede the Lick abundances for the three re-observed galaxies (KISSR 85, KISSR 666, KISSB 23), are also tabulated.  The results from the different datasets are in good agreement.

\section{The L-Z Relationship}
\subsection{\ion{H}{2} Galaxies vs. Dwarf Irregulars}

To investigate the relationship between luminosity and metallicity for the population of \ion{H}{2} galaxies, we use the abundance results presented here and in Paper II, along with additional data from the literature.  As noted in the introduction, the L-Z relation for more quiescent dwarf irregular galaxies has already been carefully quantified by exclusively using $T_e$-determined abundances and accurate photometry (e.g. Skillman et al. 1989, Richer \& McCall 1995, H. Lee et al. 2003).  However a similar treatment for \ion{H}{2} galaxies, so named because their light is dominated by star-formation occurring in compact regions (Melnick et al. 1985), is not available.  In fact, previous investigations of the L-Z relation in the low-luminosity regime have typically excluded galaxies that are strongly star-forming since these objects are thought to obscure any relationship that might exist.  An implicit assumption here is that the fundamental relationship is between mass and metallicity (e.g. Lequeux et al. 1979, Tremonti et al. 2004), and that the overwhelming contribution of light from starbursting regions causes the observed B luminosity to be a poor indicator of stellar mass.

From previously published work, we select \ion{H}{2} galaxies with abundances from Kobulnicky \& Skillman (1996) and Izotov \& Thuan (1999) that have B photometry in either Salzer et al. (1989) or Gil de Paz et al. (2003).  In the analysis that follows, {\it we only use abundances that have been determined directly using the $T_e$ method}.  Magnitudes are corrected for Galactic extinction and we adopt an $H_o$ of 75 km s$^{-1}$ Mpc$^{-1}$.  All of these data are compiled in Table~\ref{tab4}.  Note that the absolute magnitudes given in Table~\ref{tab4} for the KISS ELGs supersede those computed using the objective-prism redshifts that are published in the original survey papers (Salzer et al. 2001, Salzer et al. 2002, Gronwall et al. 2004).  Redshifts based on long-slit spectral data are a great deal more accurate ($\sigma=$ 15-30 km s$^{-1}$) than those estimated from the coarse objective-prism spectra ($\sigma\sim$ 850 km s$^{-1}$).  A conservative value for the typical uncertainty in the KISS absolute magnitudes would then be 0.3 mag.  This mainly reflects the error in the recessional velocity-based distance, since the photometric uncertainties are small, ranging from 0.03 mags for the brighter objects in the sample, to $\sim$ 0.1 for galaxies with B$\sim$20 (Salzer et al. 2001).

\begin{figure}
\epsscale{1.0}
\plotone{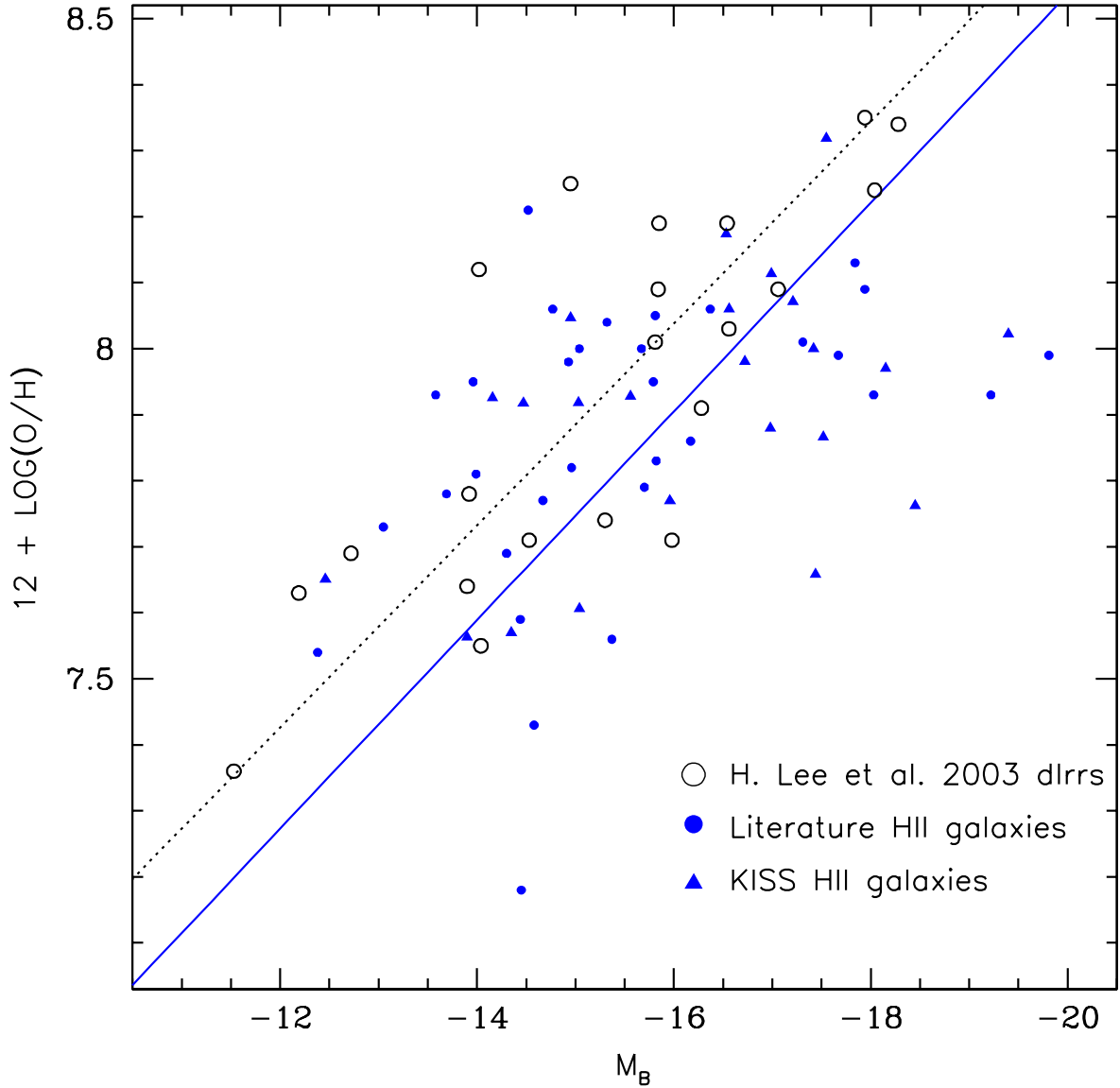}
\figcaption{Comparison of \ion{H}{2} galaxies (solid symbols) and more quiescent dwarf irregular galaxies (open circles) in the L-Z plane.  The solid line is a least-squares bisector fit to the entire \ion{H}{2} galaxy sample while the dotted line is the fit to the dIrr galaxies calculated by H. Lee et al. 2003.  \ion{H}{2} galaxies and dIrrs follow very similar luminosity metallicity relationships, with the primary difference being a small zero-point offset such that the \ion{H}{2} galaxies are shifted to higher luminosities, or equivalently, to lower abundances. \label{lz}}
\end{figure}

The positions of the \ion{H}{2} galaxies in the L-Z plane are shown in Figure~\ref{lz} as filled symbols, where the triangles are KISS ELGs (N=23) and the circles represent the galaxies from the literature (N=31).  Both samples occupy the same areas of parameter space, and show the familiar trend of increasing oxygen abundance with increasing luminosity.  To fit the data we compute the bisector of an initial ordinary least-squares fit where $M_B$ and $Z$ are taken as the independent and dependent variables respectively, and a second fit where the variables are interchanged.  The results of the fits (given in Table~\ref{fits}) show that the KISS and literature \ion{H}{2} galaxy samples, whether combined or considered separately, define the same L-Z relationship to within the errors.  The fit to the composite \ion{H}{2} galaxy sample is shown in Figure~\ref{lz} by the solid line.

In Figure~\ref{lz}, we also overplot nearby ``normal'' field dwarf irregulars (i.e. those that are not undergoing a massive, concentrated, burst of star-formation at the present epoch) from H. Lee et al. (2003) as open circles.  This most recent determination of the L-Z relation for dwarf irregulars is essentially an update of the work published by Richer \& McCall (1995), where H. Lee et al. (2003) follow their protocol of only including galaxies with $T_e$ based oxygen abundances and distances determined using stellar indicators, and find that the addition of new data does not significantly change the earlier result.  Both the Richer \& McCall (1995) and H. Lee et al. (2003) results are also consistent with the relationship found by Skillman, Kennicutt \& Hodge (1989).

Figure~\ref{lz} demonstrates that the populations of \ion{H}{2} galaxies and more quiescent dIrrs define very similar luminosity-metallicity relationships, albeit with a small shift such that \ion{H}{2} galaxies tend towards larger luminosities at a given oxygen abundance.  Comparison of the formal fit for the dIrrs computed in H. Lee et al. (2003) (given in Table~\ref{fits} and shown in Figure~\ref{lz} as a dotted line) and our fit for the total \ion{H}{2} galaxy sample confirms that a subtle zero-point shift is detectable.  It would appear as though the difference is not significant because of the $\sim$0.5 dex formal error on the zero-point.  However, this is due to the covariance of the slope and the zero-point -- a small error in the slope results in a large error in the intercept due to the large lever arm of the points relative to the y-axis.  If we instead perform a constrained least-squares fit with the slope fixed at -0.153, which is the value reported by both Skillman et al. (1989) and H. Lee et al. (2003) for the dIrrs, the resulting zero-point for the total \ion{H}{2} galaxy sample is 5.46 $\pm$ 0.03, and the error in the zero-point for the H. Lee et al. (2003) dIrrs becomes $\pm$ 0.04.  These statistics suggest that there is a luminosity shift of 0.8 B magnitudes between the correlations followed by the two populations which is significant at the 3 to 4 $\sigma$ level.

\subsection{Physical Interpretation of Shifts in the L-Z Plane}
How can this difference in the correlations be physically understood?  It seems reasonable to attribute the zero-point offset to a systematic luminosity shift in the populations, particularly if the fundamental correlation is between stellar mass and metallicity (Tremonti et al. 2004).  Since \ion{H}{2} galaxies have B luminosities that are less representative of the total stellar mass because of the large contribution of light from the starbursting regions, we can possibly expect these galaxies to show a shift to higher luminosities when compared with other galaxies that are not forming stars as vigorously.  As discussed above, we see such a shift in Figure~\ref{lz}. However, we can go one step further and interpret the offset as indicative of the {\it amount of the average luminosity enhancement} in the \ion{H}{2} galaxy population.  That is, the results of our analysis suggest a scenario where the typical \ion{H}{2} galaxy is in a state of moderate brightening in which the starburst component is contributing about 0.8 magnitude of additional light in the B-band.  This conclusion is consistent with independent studies that have used surface brightness decompositions to separate star-forming components from the underlying ``host'' in low-luminosity \ion{H}{2} galaxies and find that the star-formation regions elevate the host luminosity by an average of 0.75 B magnitudes (Salzer \& Norton 1998, Papaderos et al. 1996 and references therein).
 
A systematic deviation of \ion{H}{2} galaxies from the established dIrr relation has also been previously noted in passing by authors who investigate the L-Z relation over a larger luminosity range for galaxies at higher redshift and use local \ion{H}{2} galaxies from the literature and the present day L-Z relationship for comparison (e.g. Kobulnicky \& Zaritsky 1999, Contini et al. 2002).  These studies also attribute this deviation to the brightening caused by the typical starburst in an \ion{H}{2} galaxy, but they show much larger average offsets of 2 to 3 magnitudes.  These results can be reconciled with our smaller quoted offset by noting that the low-luminosity ends of their L-Z relationships are defined with the data from Richer \& McCall (1995).  Using the more recently updated version of these data in H. Lee et al. (2003), as is done here, yields an approximate 1 magnitude shift towards higher luminosities, which reduces the apparent offsets between the two populations in those papers to 1 to 2 magnitudes.  This is more consistent with our results.  Therefore if Figure~\ref{lz} is robust, and the luminosity enhancement interpretation is correct, we may also say that prodigious 2 to 3 magnitude luminosity enhancements are atypical and do not occur in the majority of \ion{H}{2} galaxies.  In other words, low mass galaxies that have episodes of star-formation that elevate their luminosities by factors of 10 to 20 are not common in the local universe. 

Let us now turn our attention to the handful of HII galaxies which are clear outliers and do in fact deviate from the relationship by up to 3 to 4 magnitudes.  In the process of re-checking the abundances and photometry for these objects, we noticed that many of them appeared to have tidal features.  But what would cause tidally distorted HII galaxies to preferentially show strong deviations from the L-Z relationship?  If some fraction of these systems have been recently involved in an interaction with an external body, lower-metallicity or primordial gas could be introduced into the ISM which surrounds the star-forming regions.  This would dilute the ISM and effectively decrease the observed metallicity if prompt enrichment from the current star-formation event does not occur.  Dilution through infall is feasible given that low-luminosity \ion{H}{2} galaxies, in particular blue compact dwarfs, have been shown to have large reservoirs of HI gas (e.g. Lee et al. 2002) that extend far beyond their optical disks (van Zee et al. 1998, 2001).  This reservoir may contain relatively unenriched gas, which through the loss of angular momentum, can fall to the center of the system where it can build up and form stars, or be ionized by \ion{H}{2} regions which are already in place.  Thus, interactions may also cause more massive starbursts which can lead to greater luminosity enhancements.  However, infall is an additional effect that must be considered in causing objects to deviate from the L-Z relationship.  Another possibility is that the \ion{H}{2} galaxy outliers may on average have lower gas abundances due to a systematically larger outflow of metal enriched gas, but we do not consider this in any detail here.

\begin{figure}
\epsscale{1.0}
\plotone{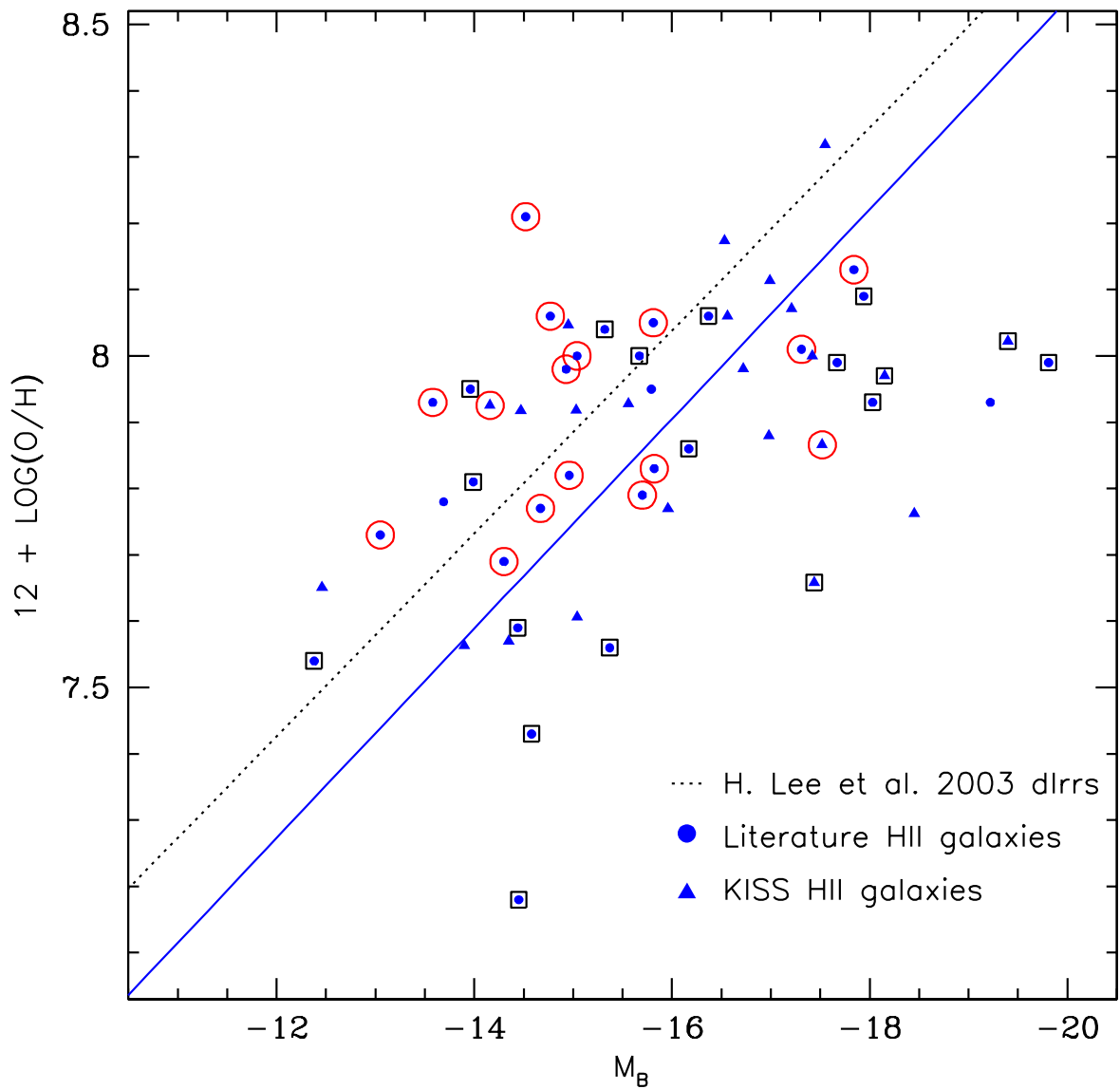}
\figcaption{The same as Figure~\ref{lz}, except that symbols for \ion{H}{2} galaxies with disturbed, irregular outer isophotes are now enclosed in open squares and symbols for the \ion{H}{2} galaxies with more regular morphologies are enclosed in open circles.  The H. Lee et al. dIrr galaxies have also been removed.  There is a slight offset in the average L-Z relations for these two populations such that the disturbed galaxies tend to be more luminous and/or more metal-poor. \label{lz_morph}}
\end{figure}

Since it is difficult to conclusively determine whether an HII galaxy has recently undergone an interaction using the images available to us on NED and our KISS survey data, we decided to test the effect of including tidally distorted objects on the L-Z relationship in the following coarse manner.  First, for the \ion{H}{2} galaxies which have photometry in the Gil de Paz (2003) atlas of BCDs, we adopt the Loose \& Thuan (1986) morphological classification given therein to indicate whether or not a system has been disturbed.  We simply consider galaxies of type iE or nE (those with regular outer elliptical isophotes) as undisturbed, and galaxies of type iI (those with irregular outer isophotes and an off-center nucleus), as disturbed.  There are also four galaxies with Gil de Paz (2003) photometry in Table~\ref{tab4} that have a classification of I0 which is given to those objects without an apparent diffuse extended component.  For these galaxies we simply use the regularity of the outermost visible isophotes to judge whether or not it is disturbed.  Finally, for the KISS ELGs and other literature \ion{H}{2} galaxies, we have used KISS imaging data and images found on NED, respectively, to divide them into these two very broad classes.  We do not attempt to make a determination for the objects which are too faint and/or too distant such that they are not well-resolved, so not all of the galaxies in Table~\ref{tab4} have been classified.  Of course, the iI-type encompasses both galaxies which are truly tidally distorted and those that simply appear irregular because star-formation has proceeded in a random way in small isolated pockets.  However, galaxies which have recently undergone interactions will nevertheless be contained in this general class.  The result of this exercise is shown in Figure~\ref{lz_morph}, where the galaxies with irregular morphologies are marked by squares and those with more regular morphologies are circled.

Figure~\ref{lz_morph} shows that although there is an offset in the average L-Z relationships for \ion{H}{2} galaxies that have irregular outer isophotes and those that do not, it is subtle.  The scatter of the correlations defined by the two populations is large.  Note however that if the points marked by squares are removed from the diagram, the correlation traced by the \ion{H}{2} galaxies becomes more consistent with the one followed by the dIrrs.  If we preform a constrained linear fit with the slope fixed at the value reported by both Skillman et al. (1989) and H. Lee et al. (2003) for the dIrrs, the sub-sample of galaxies with regular outer isophotes has a y-intercept of 5.60$\pm$0.05, which is the same as that for the dIrr sample.  On the other hand, the value of the intercept for the HII galaxy sample with more irregular outer isophotes is 5.34$\pm$0.06.  

In and of itself, this segregation between these two broad morphological groups in the L-Z plane is interesting and needs to be explained.  Whether or not the subset of truly tidally disturbed systems within the irregular class is indeed the component driving this offset remains to be established with a larger sample of unambiguously interacting systems.  A more detailed future study of this kind may also benefit from a quantitative approach to identifying tidally disturbed galaxies such as that described in Conselice et al. (2000).

To summarize, there are two competing effects that can be responsible for causing galaxy samples to systematically deviate from the L-Z relationship towards higher luminosities and/or lower metallicities.  Either the luminosities of the galaxies can be enhanced by star-formation, or the sample can have systematically lower gas-phase abundances via inflows or outflows.  We note however that if flows play any role in producing the offset between the populations of HII galaxies and dIrrs shown in Figure~\ref{lz}, the role of luminosity enhancement due to star-formation  must decrease, thereby strengthening the argument that low mass galaxies that have episodes of star-formation that increase their light production by factors of 10 to 20 are not typical.

To make more quantitative conclusions regarding the actual fraction of dwarf galaxies that undergo extreme bursts requires the analysis of complete samples with well-defined selection functions, which we do not have here.  There are also a number of tests that can be performed to help distinguish between luminosity enhancement and gas phase metal depletion/dilution as the cause of the offset between the L-Z relationships defined by the two populations.  These include studying the change in the relative offset as the relationship is examined at longer wavelengths, and checking the predictions of closed-box models against the chemical and gas properties of the galaxies in our samples.  We leave these for future work.
\subsection{Possible Consequences for Inferring Chemical Evolution from High Redshift L-Z Relations }

If the inclusion of interacting galaxies does indeed move the observed L-Z correlation towards lower metallicities and/or larger luminosities, there is an interesting implication for recent studies which attempt to measure the build-up of metals over time by investigating the L-Z relationship for galaxies at higher redshifts (Lilly et al. 2003, Kobulnicky et al. 2003).  Currently favored hierarchical models of galaxy formation predict that the fraction of interacting and merging galaxies should increase with increasing look-back time (e.g. Kauffman \& White 1993).  The well-known observational finding that galaxy samples at higher redshift contain larger fractions of blue, morphologically irregular systems substantiates these models (e.g. van den Bergh et al. 1996).  Thus, L-Z relationships determined using higher z galaxies may be shifted towards lower metallicities and/or larger luminosities when compared with the local relation not only because the samples may contain galaxies which are less chemically evolved, but also because they will contain larger fractions of interacting galaxies.  This must be taken into account when inferring abundance evolution from changes in the L-Z relationship with cosmic time.  A more rigorous determination of the effect of interacting galaxies on the local L-Z relationship would be valuable in this regard.

\section{Summary}
This paper presents the second installment of KISS ELGs with follow-up spectroscopy of high enough quality to allow for the determination of nebular abundances.  Here we have reported the results of spectroscopic observations of 14 KISS star-forming ELGs.  These galaxies are determined to be low metallicity objects with 7.61 $\leq$ 12 + log(O/H) $\leq$ 8.32  Combining the data presented here and in Paper II, there are now 23 KISS ELGs for which $T_e$ based metallicities can be computed.

By using these abundances in conjunction with other $T_e$-based measurements from the literature, we demonstrate that low-luminosity \ion{H}{2} galaxies and the more quiescent dwarf irregular galaxies follow very similar metallicity-luminosity relationships.  The primary difference is a small zero-point shift between the correlations.  We consider infall of metal-poor gas and temporary decreases in mass-to-light ratios due to vigorous star-formation in the \ion{H}{2} galaxies as possible mechanisms for producing the offset.  We argue that the offset in the L-Z relations followed by the two populations can be used as a constraint on the average brightening experienced by \ion{H}{2} galaxies, and conclude that the typical low-luminosity \ion{H}{2} galaxy is in a moderate state of luminosity enhancement and are brighter by an average of 0.8 B magnitudes at a given metallicity.  This suggests that low-mass galaxies which host prodigious star-formation episodes that elevate the luminosities of their host galaxies by 2 to 3 magnitudes are not common at the present epoch.  

We also show that there is a segregation of the \ion{H}{2} galaxy population by morphology in the L-Z plane in the sense that \ion{H}{2} galaxies with disturbed, irregular outer isophotes tend to be more luminous and/or more metal-poor than ones that are more regularly structured.  Interactions with external bodies which cause the infall of metal deficient gas or the onset of vigorous star-formation may produce this effect.  If the inclusion of interacting galaxies does indeed increase the scatter in the L-Z relation and force the observed correlation towards lower metallicities and/or larger luminosities, this must be taken into account when attempting to infer metal abundance evolution from galaxy samples at higher redshift via changes in the L-Z relation since these samples will contain larger fractions of interacting galaxies. 

The data presented here is further used in the next paper in the KISS abundance series (Salzer et al. 2004; Paper IV) to improve the calibration of our empirical strong-lined metallicity estimator, to update the KISS ELG L-Z relationship over the full range of luminosity in the B-band, and further, to determine it in near-infrared J, H and K-bands.  If the fundamental relationship is between baryonic mass and metallicity, and the luminosity of the galaxy is being used as an observable indicator for mass, then it is desirable to examine the correlation in the near-infrared, since the light at these wavelengths is less affected by recent star-formation and dust.

\acknowledgments
We have greatly benefited from discussions with Rob Kennicutt, Christy Tremonti, Don Garnett and John Moustakas.  JCL acknowledges financial support from NSF grants AST-9617826 and AST-0307386, and a UA-NASA Space Grant Graduate Fellowship.  We also acknowledge financial support for the KISS project from an NSF Presidential Faculty Award to JJS (NSF-AST-9553020), as well as continued support for our ongoing follow-up spectroscopy campaign (NSF-AST-0071114).  We thank the many KISS team members who have participated in the spectroscopic follow-up observations during the past several years, particularly Caryl Gronwall, Drew Phillips, Gary Wegner, Anna Jangren, Jessica Werk, Laura Chomiuk, and Robin Ciardullo.




\clearpage
\begin{deluxetable}{ccccrcccccl}
\tabletypesize{\scriptsize}
\rotate
\tablecaption{KISS-MMT Spectroscopic Sample: General Properties \label{tab1}}
\tablewidth{0pt}
\tablehead{
\colhead{KISSR} &
\colhead{KISSB} &
\colhead{R.A.} &
\colhead{Decl.}  & 
\colhead{$cz$} & 
\colhead{$m_B$} &  
\colhead{$B-V$} & 
\colhead{$M_B$}&
\colhead{$SFR$}&
\colhead{t$_{exp}$}&
\colhead{Other Names}\\
\colhead{ID} &
\colhead{ID} &
\colhead{(J2000)} &
\colhead{(J2000)} &
\colhead{(km s$^{-1}$)} & 
&
&
&
\colhead{(M$_{\sun}$ yr$^{-1}$)}&
\colhead{(sec)}\\
\colhead{(1)}&\colhead{(2)}&\colhead{(3)}&\colhead{(4)}&\colhead{(5)}&\colhead{(6)}&\colhead{(7)}&\colhead{(8)}&\colhead{(9)}&\colhead{(10)}&\colhead{(11)}
}
\startdata

\nodata &23\tablenotemark{*}   &09 40 12.7 &29 35 29&  473         &16.24 &0.22 &-12.46  &0.0004 &900&KUG 0937+298, PGC 027591\\
\nodata &61  &11 19 31.9 &29 09 29&24486                           &19.16 &0.55  &-18.45  &3.6954 &600&\nodata\\
\nodata &86  &12 10 16.9 &28 45 27& 6508                           &18.13 &0.36  &-16.56  &0.1036 &900&CG 159,  LEDA 142893\\
73      &98  &12 31 57.2 &29 42 46&  632                           &15.47 &0.33  &-14.16  &0.0011 &600&\nodata\\
85\tablenotemark{*}  &\nodata &12 37 18.5 &29 14 55& 7021          &19.83 &0.11  &-15.04  &0.0525 &600&\nodata\\
87      &104 &12 38 08.1 &29 10 50&19780                           &19.59 &0.66  &-17.55  &0.7661 &900&\nodata\\
116     &107 &12 46 38.7 &29 27 37& 9569                           &18.34 &0.42  &-17.21  &0.2828 &900&WAS 63, CG 195, LEDA 140006\\
286     &133 &13 26 25.1 &29 10 32& 5174                           &17.68 &0.54  &-16.53  &0.3692 &900&WAS 70, UCM 1324+2926, LEDA 140180\\
310     &136 &13 33 45.3 &28 45 12&10660                           &18.81 &0.59  &-16.98  &0.7708 &600&WAS 74, UCM 1331+2900\\
311     &137 &13 33 55.8 &29 21 52& 9626                           &18.84 &0.61  &-16.72  &0.1901 &600&\nodata\\
\nodata &171 &14 33 50.0 &29 34 09& 6484                           &17.73 &0.36  &-16.99  &0.1598 &900&CG 1240, UCM 1431+2947\\
\nodata &175 &14 44 00.1 &29 15 55&13733                           &16.95 &0.42  &-19.40  &2.8718 &600&CG 1258\\
666\tablenotemark{*}&186\tablenotemark{*}&15 15 42.5&29 01 40&10009&19.71 &0.39  &-15.96  &0.2598 &900&\nodata\\
814 &199 &15 35 56.0 &29 44 20&15328                               &18.45 &0.71  &-18.15  &0.7063 &900&\nodata\\
\enddata

\tablecomments{Distance dependent quantities assume $H_o = $ 75 km s$^{-1}$ Mpc$^{-1}$ and are based on the follow-up spectral redshifts rather than on the coarse objective-prism redshifts.  Col.(1): ID from the first red survey list (30\degr Red Survey; Salzer et al. 2001), where selection is based on H$\alpha$.   Col.(2): ID from the blue survey list (30\degr Blue Survey; Salzer et al. 2002), where selection is based on [OIII]$\lambda$5007.  Col.(5): Calculated from the average of the redshifts measured from all of the strong emission lines from the object's MMT spectrum.  Typical formal errors in $z$ range from 0.00005 to 0.0001 (15 to 30 km s$^{-1}$)    Col.(6): Taken from Salzer et al. 2001, Salzer et al. 2002 and corrected for Galactic extinction using Schlegel et al. 1998. Col.(7): Computed from apparent magnitudes given in Salzer et al. 2001, Salzer et al. 2002 and corrected for Galactic extinction using Schlegel et al. 1998.  Col.(9): Based on objective-prism fluxes.  While these values are valid from a statistical point of view (i.e., for the entire KISS ensemble), there are likely to be significant uncertainties for individual objects.  We show these values to simply illustrate the range of levels of star formation present in the current sample.  Col.(10): Total exposure time.}

\tablenotetext{*}{Repeat observation of target from Paper II.}
\end{deluxetable}

\clearpage
\renewcommand{\arraystretch}{1} 
\begin{deluxetable}{lrrrrrrrr}
\tabletypesize{\scriptsize}
\tablewidth{0pt}
\tablecaption{Corrected Line Ratios with respect to H$\beta$ \label{tab2}}
\tablehead{
&
& 
\colhead{KISSB 23}&
\colhead{KISSB 61}& 
\colhead{KISSB 86}& 
\colhead{KISSR 73}& 
\colhead{KISSR 85}& 
\colhead{KISSR 87}& 
\colhead{KISSR 116} 
}
\startdata
\\[-2ex]
\multicolumn{2}{c}{cH$\beta$}  &0.000(0.053) & 0.000(0.053) & 0.000(0.052) & 0.047(0.071) & 0.002(0.074) & 0.023(0.052) & 0.011(0.052) \\ 
\\[-2ex]
\hline
 $[$O II$]$ B  &       3727 & 247.4(6.7) &   138.7(3.9) &   214.8(5.7) &   261.5(18.2) &   117.3(8.8) &   189.5(8.8) &   35.1(1.5) \\
 H 10  &       3798 &  &  &  &  &  &  &  \\ 
 H 9  &       3836 &  &   2.4(0.3) &  &  &  &   3.0(0.2) &  \\ 
 $[$Ne III$]$  &       3869 & 11.4(0.5) &   37.6(1.2) &   38.7(1.1) &   29.6(2.0) &  &   31.2(1.3) &   35.1(1.5) \\
 He I+H 8  &       3889 & 10.5(0.5) &   15.3(0.6) &   11.7(0.4) &   16.8(1.2) &   7.4(1.4) &   14.3(0.6) &   9.5(0.4) \\
 $[$Ne III$]$  &       3970 & 13.1(0.5) &   22.9(0.8) &   14.8(0.5) &   23.2(1.5) &   10.2(1.4) &   20.4(0.8) &   16.4(0.7) \\
 H $\delta$  &       4102 & 17.3(0.6) &   20.2(0.7) &   14.9(0.5) &   25.1(1.5) &   21.6(1.9) &   22.9(0.9) &   15.9(0.6) \\
 He I  &       4026 &  &  &  &  &  &  &  \\ 
 $[$S II$]$ B  &       4073 &  &  &  &  &  &  &  \\ 
 H $\gamma$ &       4340 & 40.8(1.2) &   44.1(1.3) &   35.9(1.0) &   47.4(2.5) &   44.4(2.8) &   41.8(1.3) &   38.1(1.2) \\
 $[$O III$]$  &       4363& 2.6(0.4) &   9.5(0.4) &   5.5(0.3) &   5.6(0.5) &   6.6(1.3) &   2.8(0.2) &   4.9(0.2) \\
 He I  &       4472 & 1.6(0.4) &  &   2.0(0.3) &   2.9(0.4) &  &   3.9(0.2) &   3.1(0.2) \\
 He II  &       4687 &  &  &  &  &  &  &  \\ 
 H $\beta$  &       4861 & 100.0(2.8) &   100.0(2.8) &   100.0(2.7) &   100.0(4.6) &   100.0(5.1) &   100.0(2.7) &   100.0(2.6) \\
 He I  &       4922 &  &  &  &  &  &  &  \\ 
 $[$O III$]$  &       4959 & 46.6(1.4) &   150.6(4.2) &   150.6(4.0) &   121.8(5.6) &   100.4(5.1) &   148.8(4.0) &   146.9(3.9) \\
 $[$O III$]$  &       5007 & 140.2(3.8) &   449.2(12.2) &   454.3(12.0) &   362.4(16.7) &   304.3(14.8) &   447.0(11.9) &   437.8(11.6) \\
 $[$N I$]$  &       5199 &  &  &  &  &  &  &  \\ 
 He I  &       5876 & &   9.6(0.5) &   10.6(0.4) &   10.8(0.7) &   10.7(1.4) &   12.2(0.5) &   10.7(0.4) \\
 $[$O I$]$  &       6300 & &   5.6(0.4) &   5.9(0.3) &  &  &   3.4(0.3) &   4.5(0.3) \\
 $[$S III$]$  &       6312 & &   3.0(0.3) &   1.1(0.2) &  &  &   0.5(0.2) &   0.8(0.2) \\
 $[$O I$]$  &       6364 &  &  &  &  &  &   0.5(0.2) &  \\ 
 $[$N II$]$  &       6548 & &  &   4.1(0.3) &   3.1(0.4) &  &   7.5(0.4) &   6.1(0.4) \\
 H $\alpha$  &       6563 & 279.3(7.6) &   274.9(7.6) &   282.2(7.5) &   280.7(20.0) &   278.3(20.9) &   286.3(13.8) &   282.7(13.6) \\
 $[$N II$]$  &       6583 & 9.9(0.5) &   1.9(0.3) &   11.0(0.4) &   8.4(0.7) &   1.0(1.2) &   22.7(1.2) &   18.0(0.9) \\
 He I  &       6678 & &   2.4(0.3) &   2.2(0.3) &   2.8(0.4) &  &   3.2(0.3) &   2.0(0.2) \\
 $[$S II$]$  &       6717 & 24.6(0.8) &   13.5(0.6) &   23.8(0.7) &   18.3(1.5) &   7.8(1.4) &   18.9(1.0) &   19.9(1.1) \\
 $[$S II$]$  &       6731 & 17.3(0.7) &   9.0(0.5) &   16.2(0.5) &   13.1(1.1) &   11.2(1.6) &   14.0(0.8) &   13.8(0.8) \\
 He I  &       7065 &  &  &   3.0(0.3) &  &  &   2.0(0.2) &   2.1(0.2) \\
 $[$Ar III$]$  &       7136 & &   3.3(0.4) &   6.8(0.3) &   5.7(0.6) &  &   5.8(0.4) &   7.8(0.5) \\
 $[$O II$]$  &       7319 &  &  &  &  &  &  &   1.7(0.2) \\
  $[$O II$]$  &       7330 &  &  &  &  &  &  &   1.2(0.2) \\
\tableline
\enddata
\end{deluxetable}

\renewcommand{\arraystretch}{1} 
\begin{deluxetable}{lrrrrrrrr}
\tabletypesize{\scriptsize}
\tablenum{2}
\tablewidth{0pt}
\tablecaption{$-$ Continued}
\tablehead{
& 
& 
\colhead{KISSR 286} & 
\colhead{KISSR 310} & 
\colhead{KISSR 311} & 
\colhead{KISSB 171} & 
\colhead{KISSB 175} & 
\colhead{KISSR 666} & 
\colhead{KISSR 814} 
}
\startdata
\\[-2ex]
\multicolumn{2}{c}{cH$\beta$}&0.223(0.052) & 0.120(0.070) & 0.150(0.071) & 0.000(0.052) & 0.121(0.052) & 0.238(0.071) & 0.158(0.052) \\
 \\[-2ex]
\hline

 $[$O II$]$ B  &       3727 & 205.3(9.4) &   95.5(6.5) &   238.6(16.4) &   183.1(4.8) &   179.0(8.2) &   13.7(1.2) &   181.4(8.4) \\
 H 10  &       3798 & &   3.8(0.3) &  &   3.0(0.2) &   4.0(0.2) &  &   3.1(0.2) \\
 H 9  &       3836 & 4.0(0.2) &  &  &   3.9(0.2) &   5.6(0.3) &  &   5.3(0.3) \\
 $[$Ne III$]$  &       3869 & 
  31.7(1.3) &   53.0(3.3) &   45.5(2.9) &   37.0(1.0) &   50.0(2.1) &   36.7(2.5) &   43.3(1.8) \\
 He I+H 8  &       3889 & 
  12.6(0.5) &   18.1(1.1) &   16.8(1.1) &   14.2(0.4) &   15.5(0.7) &   14.5(1.1) &   15.0(0.7) \\
 $[$Ne III$]$  &       3970 & 
  19.9(0.8) &   30.4(1.8) &   25.3(1.6) &   23.2(0.6) &   27.5(1.1) &   20.3(1.4) &   28.0(1.1) \\
 H$\delta$  &       4102 & 
  19.2(0.7) &   26.1(1.5) &   27.1(1.6) &   21.2(0.6) &   23.6(0.9) &   17.6(1.2) &   25.7(1.0) \\
 He I  &       4026 &  &   1.3(0.1) &  &  &  &  &  \\ 
 $[$S II$]$ B  &       4072 &  &   1.4(0.1) &  &  &  &  &  \\ 
 H$\gamma$  &       4340 & 
  41.1(1.3) &   47.7(2.4) &   49.2(2.5) &   43.5(1.2) &   47.8(1.5) &   39.5(2.1) &   46.8(1.5) \\
 $[$O III$]$  &       4363& 
  3.7(0.2) &   12.9(0.7) &   7.9(0.5) &   5.2(0.2) &   8.5(0.3) &   16.4(1.0) &   8.0(0.3) \\
 He I  &       4472& 
  3.5(0.1) &   3.8(0.2) &   4.1(0.4) &   3.6(0.2) &   3.8(0.2) &  &   2.4(0.2) \\
 He II  &       4687 &  &   2.2(0.1) &  &  &  &  &  \\ 
 H $\beta$  &       4861 & 
  100.0(2.6) &   100.0(4.5) &   100.0(4.6) &   100.0(2.6) &   100.0(2.6) &   100.0(4.7) &   100.0(2.7) \\
 He I  &       4922 &  &   0.8(0.1) &  &   1.2(0.1) &   1.0(0.1) &  &  \\ 
 $[$O III$]$  &       4959& 
  142.0(3.7) &   214.3(9.7) &   164.2(7.5) &   161.0(4.2) &   190.9(5.0) &   228.7(10.6) &   171.2(4.6) \\
 $[$O III$]$  &       5007 & 
  426.3(11.2) &   644.0(29.1) &   493.5(22.5) &   485.7(12.7) &   569.4(15.0) &   692.5(31.9) &   512.9(13.7) \\
 $[$N I$]$  &       5199 &   1.1(0.1) &  &  &  &  &  &  \\ 
 He I  &       5876 & 
  11.1(0.4) &   9.6(0.6) &   10.1(0.7) &   10.1(0.3) &   10.2(0.4) &   9.8(0.7) &   11.1(0.5) \\
 $[$O I$]$  &       6300 & 
  4.7(0.2) &   2.0(0.2) &   5.3(0.4) &   4.3(0.2) &   4.1(0.2) &  &   4.0(0.2) \\
 $[$S III$]$  &       6312 & 
  1.0(0.1) &   1.6(0.1) &   1.9(0.3) &   1.4(0.1) &   1.0(0.1) &  &   1.7(0.2) \\
 $[$O I$]$  &       6364 & 
  1.1(0.1) &   0.7(0.1) &  &   1.0(0.1) &   1.3(0.1) &  &  \\ 
 $[$N II$]$  &       6548 & 
  5.1(0.3) &   1.1(0.1) &   2.9(0.3) &   4.0(0.2) &   3.0(0.2) &  &   6.1(0.3) \\
 H$\alpha$  &       6563 & 
  284.5(13.6) &   278.9(19.6) &   280.5(19.9) &   283.1(7.4) &   280.8(13.5) &   277.8(20.0) &   280.5(13.5) \\
 $[$N II$]$  &       6583 & 
  15.3(0.8) &   3.2(0.2) &   8.1(0.6) &   10.9(0.3) &   8.6(0.4) &   2.4(0.4) &   13.6(0.7) \\
 He I  &       6678 & 
  3.0(0.2) &   3.1(0.3) &   2.6(0.3) &   3.1(0.2) &   2.3(0.2) &   2.4(0.4) &   2.1(0.2) \\
 $[$S II$]$  &       6717 & 
  21.3(1.1) &   7.3(0.6) &   16.8(1.3) &   19.9(0.6) &   14.9(0.8) &   3.4(0.4) &   18.8(1.0) \\
 $[$S II$]$  &       6731 & 
  15.3(0.8) &   5.3(0.4) &   12.2(1.0) &   13.8(0.4) &   11.8(0.6) &   2.1(0.4) &   14.1(0.8) \\
 He I  &       7065 & 
  1.7(0.1) &  &  &   2.1(0.1) &   2.4(0.2) &  &   2.8(0.2) \\
 $[$Ar III$]$  &       7136 & 
  6.9(0.4) &  &  &   7.9(0.3) &   5.7(0.3) &  &   6.6(0.4) \\
 $[$O II$]$  &       7319 & 
  2.1(0.1) &  &  &   2.3(0.1) &   2.3(0.2) &  &  \\ 
 $[$O II$]$  &       7330 & 
  1.9(0.1) &  &  &   1.6(0.1) &   1.8(0.2) &  &  \\ 
\tableline 
 
\enddata
\end{deluxetable}

\begin{deluxetable}{lrrrrrrr}
\renewcommand{\arraystretch}{1.2}
\tabletypesize{\scriptsize}
\tablewidth{0pt}
\tablecaption{Derived Metal Abundances \label{tab3}}
\tablehead{
                & 
\colhead{KISSB 23} & 
\colhead{KISSB 61} & 
\colhead{KISSB 86} & 
\colhead{KISSR 73} & 
\colhead{KISSR 85} & 
\colhead{KISSR 87} & 
\colhead{KISSR 116}
}
\startdata
N$_e$ (cm$^{-3}$) & \multicolumn{1}{c}{100}&\multicolumn{1}{c}{100}&\multicolumn{1}{c}{100}&\multicolumn{1}{c}{35}&\multicolumn{1}{c}{100}&\multicolumn{1}{c}{71}&\multicolumn{1}{c}{100}\\
T$_e$ [OII] (K) &    13530(492)&13897(205)&12445(189)&13035(350)&13984(716)&11068(181)&12256(153)\\
T$_e$ [OIII] (K) &    14745(1221)&15647(529)&12390(382)&13619(788)&15867(1975)&9931(295)&12022(299)\\

O$^+$/H$^+$ ($\times10^{-5}$) & 2.92(0.33)&1.50(0.07)&3.40(0.18)&3.45(0.30)&1.24(0.18)&4.64(0.29)&3.23(0.15)\\
O$^{++}$/H$^+$ ($\times10^{-5}$) & 1.55(0.31)&4.29(0.44)&8.09(0.88)&4.97(0.89)&2.80(0.77)&16.18(1.94)&8.55(0.81)\\
O/H ($\times 10^{-5}$) & 4.47(0.45)&5.79(0.45)&11.49(0.90)&8.42(0.94)&4.04(0.79)&20.82(1.96)&11.78(0.83)\\

N$^+$/H$^+$ ($\times 10^{-6}$) & 0.95(0.07)&\nodata&1.30(0.04)&0.89(0.05)&\nodata&3.42(0.13)&2.15(0.06)\\
N/H  ($\times 10^{-6}$) & 1.46(0.25)&\nodata&4.40(0.47)&2.18(0.38)&\nodata&15.38(2.01)&7.84(0.79)\\

Ne$^{++}$/H$^+$ ($\times 10^{-6}$) & 3.27(0.66)&9.11(0.80)&19.26(1.87)&10.82(1.72)&\nodata & 35.56(3.99)&19.38(1.58)\\
Ne/H ($\times 10^{-6}$) & 9.44(2.90)&12.29(1.95)&27.34(4.58)&18.34(5.00)&\nodata & 45.75(8.87)&26.70(4.00)\\

S$^+$/H$^+$     ($\times10^{-7} $) & 5.02(0.33)&2.56(0.07)&5.66(0.17)&3.98(0.20)&2.14(0.19)&5.95(0.21)&4.92(0.12)\\
S$^{++}$/H$^+ $     ($\times10^{-7} $) &\nodata  &\nodata &\nodata &\nodata &\nodata  &\nodata &\nodata \\
S/H     ($\times10^{-6} $) &\nodata  &\nodata &\nodata &\nodata &\nodata  &\nodata &\nodata \\

Ar$^{++}$/H$^+ $     ($\times10^{-7} $) &\nodata  & 1.24(0.07)&3.93(0.23)&2.74(0.27)&\nodata & 5.43(0.36)&4.79(0.23)\\
Ar/H     ($\times10^{-7} $) &\nodata  & 2.09(0.29)&6.27(0.54)&3.99(0.67)&\nodata & 9.86(1.05)&7.89(0.70)\\

12 + log(O/H)$_{T_e}$ 	 & 7.65(0.04)&7.76(0.03)&8.06(0.03)&7.93(0.05)&7.61(0.09)&8.32(0.04)&8.07(0.03)\\


\tableline\\[-1.5ex]
& 
\multicolumn{1}{c}{KISSR 286} & 
\multicolumn{1}{c}{KISSR 310} & 
\multicolumn{1}{c}{KISSR 311} & 
\multicolumn{1}{c}{KISSB 171} & 
\multicolumn{1}{c}{KISSB 175} & 
\multicolumn{1}{c}{KISSR 666}& 
\multicolumn{1}{c}{KISSR 814}\\[1ex]

\tableline\\[-1.5ex]

N$_e$ (cm$^{-3}$) &  \multicolumn{1}{c}{100}&\multicolumn{1}{c}{23}&\multicolumn{1}{c}{83}&\multicolumn{1}{c}{100}&\multicolumn{1}{c}{169}&\multicolumn{1}{c}{100}&\multicolumn{1}{c}{83}\\
T$_e$ [OII] (K) &    11707(167)&13746(289)&13133(288)&12162(147)&12948(158)&14248(335)&13063(168)\\
T$_e$ [OIII] (K) &    11010(298)&15268(733)&13834(656)&11843(280)&13430(344)&16565(935)&13681(372)\\

O$^+$/H$^+$ ($\times10^{-5}$) & 3.99(0.21)&1.05(0.06)&3.11(0.22)&3.09(0.14)&2.44(0.10)&0.14(0.01)&2.41(0.10)\\
O$^{++}$/H$^+$ ($\times10^{-5}$) & 10.92(1.15)&6.54(0.98)&6.46(1.01)&9.90(0.91)&8.08(0.74)&5.75(0.92)&6.92(0.65)\\
O/H    ($\times 10^{-5}$) & 14.91(1.17)&7.59(0.98)&9.57(1.03)&12.99(0.92)&10.52(0.74)&5.88(0.92)&9.33(0.66)\\

N$^+$/H$^+$ ($\times 10^{-6}$) & 2.02(0.07)&0.30(0.01)&0.84(0.04)&1.35(0.04)&0.91(0.02)&\nodata&1.52(0.04)\\
 N/H  ($\times 10^{-6}$) & 7.55(0.85)&2.16(0.35)&2.59(0.40)&5.67(0.52)&3.94(0.39)&\nodata&5.90(0.60)\\

 Ne$^{++}$/H$^+$ ($\times 10^{-6}$) & 24.07(2.28)&13.76(1.70)&15.83(2.08)&21.53(1.69)&19.09(1.46)&7.61(1.02)&15.60(1.24)\\
 Ne/H ($\times 10^{-6}$) & 32.87(5.50)&15.97(3.85)&23.45(5.61)&28.26(4.03)&24.86(3.59)&7.79(2.09)&21.02(3.12)\\

S$^+$/H$^+$     ($\times10^{-7} $) & 5.89(0.18)&1.44(0.05)&3.59(0.15)&5.00(0.12)&3.54(0.08)&0.60(0.02)&4.20(0.11)\\
 S$^{++}$/H$^+ $     ($\times10^{-7} $) & 16.49(1.64)&8.44(1.11)&13.70(1.91)&17.43(1.45)&7.95(0.64)&\nodata & 12.72(1.08)\\
S/H     ($\times10^{-6} $) & 3.13(0.30)&1.93(0.28)&2.28(0.33)&3.29(0.25)&1.70(0.14)&\nodata & 2.40(0.21)\\

 Ar$^{++}$/H$^+ $     ($\times10^{-7} $) & 5.11(0.29)&\nodata & \nodata & 5.00(0.23)&2.81(0.13)&\nodata & 3.14(0.15)\\
Ar/H     ($\times10^{-7} $) & 8.50(0.78)&\nodata &\nodata  & 8.78(0.61)&5.00(0.39)&\nodata & 5.32(0.46)\\

12 + log(O/H)$_{T_e}$ 	 & 8.17(0.03)&7.88(0.06)&7.98(0.05)&8.11(0.03)&8.02(0.03)&7.77(0.07)&7.97(0.03)\\

\enddata
\end{deluxetable}

\clearpage
\begin{deluxetable}{ccccc}
\tabletypesize{\footnotesize}
\tablecaption{Metal Abundance Ratios: Average Values and Dispersions\label{avg_ratio}}
\tablewidth{0pt}
\tablehead{
&
\colhead{KISS ELGs (MMT)}&
\colhead{KISS ELGs (Lick)}&
\colhead{KISS ELGs (Total)}&
\colhead{Thuan \& Izotov (1999)}
}
 
\startdata
$\langle$log(N/O)$\rangle$  &-1.38$\pm$0.16(11) &-1.42$\pm$0.14(12) &-1.39$\pm$0.14(22)  & -1.47$\pm$0.14(53)   \\ 
$\langle$log(Ne/O)$\rangle$ &-0.67$\pm$0.07(13) &-0.65$\pm$0.07(12) &-0.66$\pm$0.07(23)  & -0.72$\pm$0.06(54)   \\ 
$\langle$log(S/O)$\rangle$  &-1.64$\pm$0.08(6)  &-1.47$\pm$0.14(6)  &-1.56$\pm$0.14(12)  & -1.56$\pm$0.06(49)   \\   
$\langle$log(Ar/O)$\rangle$ &-2.28$\pm$0.08(9)  &-2.20$\pm$0.13(8)  &-2.23$\pm$0.12(16)  & -2.26$\pm$0.09(53)   \\   
\enddata 
\tablecomments{The number of measurements averaged is given in parentheses.}
\end{deluxetable}

\clearpage
\begin{deluxetable}{llcccccc}
\renewcommand{\arraystretch}{.85}     
\tabletypesize{\scriptsize}
\tablecaption{Luminosities and $T_e$ Abundances for KISS ELGs and \ion{H}{2} Galaxies from the Literature\label{tab4}}
\tablewidth{0pt}
\tablehead{
\colhead{ID 1}&
\colhead{ID 2}&
\colhead{R.A. (J2000)} &
\colhead{Decl. (J2000)}& 
\colhead{12 + log(O/H)}&
\colhead{Ref.}&
\colhead{$M_B$}&
\colhead{Ref.}\\
\colhead{(1)}&
\colhead{(2)}&
\colhead{(3)}&
\colhead{(4)}&
\colhead{(5)}&
\colhead{(6)}&
\colhead{(7)}&
\colhead{(8)}
}
 
\startdata
\multicolumn{8}{c}{KISS ELGs}\\
\tableline
 KISSB 23    &\nodata         &09 40 12.7&$+$29 35 29  & 7.65$\pm$0.04  &                5&   -12.46 &9  \\
 KISSB 61    &\nodata         &11 19 31.9&$+$29 09 29  & 7.76$\pm$0.03  &                5&   -18.45 &9  \\
 KISSR 1194  &\nodata         &12 03 23.0&$+$43 44 39  & 7.92$\pm$0.04  &                6&   -14.47 &2  \\
 KISSB 86    &\nodata         &12 10 16.9&$+$28 45 27  & 8.06$\pm$0.03  &                5&   -16.56 &9  \\
 KISSR 49    &\nodata         &12 24 35.2&$+$29 27 32  & 8.00$\pm$0.06  &                6&   -17.42 &8  \\
 KISSR 73    &\nodata         &12 31 57.2&$+$29 42 46  & 7.93$\pm$0.05  &                5&   -14.16 &8  \\
 KISSR 85    &\nodata         &12 37 18.5&$+$29 14 55  & 7.61$\pm$0.09  &                5&   -15.04 &8  \\
 KISSR 87    &\nodata         &12 38 08.1&$+$29 10 50  & 8.32$\pm$0.04  &                5&   -17.55 &8  \\
 KISSR 116   &\nodata         &12 46 38.7&$+$29 27 37  & 8.07$\pm$0.03  &                5&   -17.21 &8  \\
 KISSR 1490  &\nodata         &13 13 16.4&$+$44 02 29  & 7.56$\pm$0.07  &                6&   -13.90 &2  \\
 KISSR 286   &\nodata         &13 26 25.1&$+$29 10 32  & 8.17$\pm$0.03  &                5&   -16.53 &8  \\
 KISSR 310   &\nodata         &13 33 45.3&$+$28 45 12  & 7.88$\pm$0.06  &                5&   -16.98 &8  \\
 KISSR 311   &\nodata         &13 33 55.8&$+$29 21 52  & 7.98$\pm$0.05  &                5&   -16.72 &8  \\
 KISSR 396   &\nodata         &13 54 54.4&$+$29 27 46  & 7.92$\pm$0.04  &                6&   -15.03 &8  \\
 KISSR 1752  &\nodata         &14 17 01.4&$+$43 30 04  & 7.57$\pm$0.04  &                6&   -14.35 &2  \\
 KISSR 1778  &\nodata         &14 25 44.1&$+$43 51 48  & 7.93$\pm$0.08  &                6&   -15.56 &2  \\
 KISSB 171   &\nodata         &14 33 50.0&$+$29 34 09  & 8.11$\pm$0.03  &                5&   -16.99 &9  \\
 KISSR 1845  &\nodata         &14 42 16.0&$+$42 49 49  & 8.05$\pm$0.04  &                6&   -14.95 &2  \\
 KISSB 175   &\nodata         &14 44 00.1&$+$29 15 55  & 8.02$\pm$0.03  &                5&   -19.40 &9  \\
 KISSR 666   &\nodata         &15 15 42.5&$+$29 01 40  & 7.77$\pm$0.07  &                5&   -15.96 &8  \\
 KISSR 675   &\nodata         &15 17 17.4&$+$29 24 25  & 7.87$\pm$0.09  &                6&   -17.52 &8  \\
 KISSR 814   &\nodata         &15 35 56.0&$+$29 44 20  & 7.97$\pm$0.03  &                5&   -18.15 &8  \\ 
 KISSR 1013  &\nodata         &16 16 39.0&$+$29 03 33  & 7.66$\pm$0.05  &                6&   -17.44 &8  \\
\tableline
\multicolumn{8}{c}{Additional Literature \ion{H}{2} Galaxies}\\
\tableline
 0218+003    & UM 420         &02 20 54.5&$+$00 33 24  & 7.93$\pm$0.05  &                3&   -19.22 &7  \\
 0248+042    & Mrk 600        &02 51 04.6&$+$04 27 14  & 7.83$\pm$0.01  &                3&   -15.82 &1  \\
 II Zw 40    &\nodata         &05 55 42.6&$+$03 23 32  & 8.09$\pm$0.01  &                4&   -17.94 &1  \\
 0635+756    & Mrk 5          &06 42 15.5&$+$75 37 33  & 8.04$\pm$0.04  &                3&   -15.32 &1  \\
 0832+699    & UGC 4483       &08 37 03.0&$+$69 46 40  & 7.54$\pm$0.01  &                3&   -12.38 &1  \\
 0917+527    & Mrk 1416       &09 20 56.1&$+$52 34 04  & 7.86$\pm$0.02  &                3&   -16.17 &1  \\
 0930+554    & I Zw 18        &09 34 02.0&$+$55 14 28  & 7.18$\pm$0.01  &                3&   -14.45 &1  \\
 0940+544N   & SBS 0940+544   &09 44 16.7&$+$54 11 33  & 7.43$\pm$0.01  &                3&   -14.58 &1  \\
 0946+558    & UGCA 184       &09 49 30.4&$+$55 34 49  & 8.00$\pm$0.01  &                3&   -15.67 &1  \\
 Tol 2       &\nodata         &09 59 21.2&$-$28 08 00  & 8.00$\pm$0.09  &                4&   -15.04 &1  \\
 1030+583    & Mrk 1434       &10 34 10.1&$+$58 03 49  & 7.79$\pm$0.01  &                3&   -15.70 &1  \\
 1102+294    & Haro 4         &11 04 58.5&$+$29 08 22  & 7.81$\pm$0.02  &                3&   -13.99 &1  \\
 1124+792    & VII Zw 403     &11 27 59.9&$+$78 59 39  & 7.69$\pm$0.01  &                3&   -14.30 &1  \\
 1135+581    & Mrk 1450       &11 38 35.6&$+$57 52 27  & 7.98$\pm$0.01  &                3&   -14.93 &1  \\
 1139+006    & UM 448         &11 42 12.4&$+$00 20 03  & 7.99$\pm$0.04  &                3&   -19.81 &7  \\
 1148-020    & UM 461         &11 51 33.0&$-$02 22 23  & 7.78$\pm$0.03  &                3&   -13.69 &7  \\ 
 1150-021    & UM 462         &11 52 37.3&$-$02 28 10  & 7.95$\pm$0.01  &                3&   -15.79 &7  \\
 Mrk 178     &\nodata         &11 33 28.9&$+$49 14 14  & 7.95$\pm$0.02  &                4&   -13.96 &1  \\
 Pox 4       &\nodata         &11 51 11.6&$-$20 36 02  & 7.93$\pm$0.06\tablenotemark{1} &4&   -18.03 &1  \\
 UM 439      &\nodata         &11 36 36.8&$+$00 48 58  & 8.05$\pm$0.02  &                4&   -15.81 &1  \\
 1223+487    & Mrk 209        &12 26 16.0&$+$48 29 37  & 7.77$\pm$0.01  &                3&   -14.67 &1  \\
 1256+351    & NGC 4861       &12 59 02.3&$+$34 51 34  & 7.99$\pm$0.01  &                3&   -17.67 &1  \\
 Tol 65      &\nodata         &12 25 46.9&$-$36 14 01  & 7.56$\pm$0.03  &                4&   -15.37 &1  \\
 1331+493    & SBS 1331+493   &13 33 22.9&$+$49 06 06  & 7.82$\pm$0.05\tablenotemark{1} &3&   -14.96 &1  \\
 Mrk 67      &\nodata         &13 41 56.5&$+$30 31 10  & 8.21$\pm$0.08  &                4&   -14.52 &1  \\
 Pox 186     &\nodata         &13 25 48.6&$-$11 36 38  & 7.73$\pm$0.02  &                4&   -13.05 &1  \\
 Tol 35      &\nodata         &13 27 06.5&$-$27 57 23  & 8.13$\pm$0.02  &                4&   -17.84 &1  \\
 1437+370    & Mrk  475       &14 39 05.4&$+$36 48 21  & 7.93$\pm$0.01  &                3&   -13.58 &1  \\
 II Zw 70    &\nodata         &14 50 56.5&$+$35 34 18  & 8.06$\pm$0.08  &                4&   -16.37 &1  \\
 1533+574    & SBS 1533+574   &15 34 13.4&$+$57 17 07  & 8.01$\pm$0.12\tablenotemark{1} &3&   -17.31 &1  \\
 1535+554    & I Zw 123       &15 37 04.2&$+$55 15 48  & 8.06$\pm$0.04  &                3&   -14.77 &1  \\

\enddata   							    				   

\tablenotetext{1}{Average abundance of multiple \ion{H}{2} regions.  Adopted error is half of the difference of the range in values.}
\tablecomments{Col.(1): Galaxy name as listed in abundance reference in column 6.  Col.(2): Galaxy name as listed in photometry reference in column 8, if different from the abundance reference galaxy name.  Col.(5): Oxygen abundance, determined directly with the electron temperature method.  Col.(6): Reference for oxygen abundance.  Col.(7): Absolute B magnitude, assuming $H_o = $ 75 km s$^{-1}$ Mpc$^{-1}$ corrected for Galactic extinction.  Col.(8): Reference for absolute B magnitude.  REFERENCES. -- (1) Gil de Paz, Madore \& Pevunova 2003; (2) Calculated from apparent magnitudes given in Gronwall et al. 2004; (3) Izotov \& Thuan 1999; (4) Kobulnicky \& Skillman 1996; (5) This Paper; (6) Melbourne et al. 2004;  (7) Salzer, MacAlpine \& Boroson 1989; (8) Calculated from apparent magnitudes given in Salzer et al. 2001; (9) Calculated from apparent magnitudes given in Salzer et al. 2002}

\end{deluxetable}

\clearpage
\begin{deluxetable}{cclc}
\tabletypesize{\footnotesize}
\tablecaption{L-Z Relationship Fits\label{fits}}
\tablewidth{0pt}
\tablehead{
\colhead{Fit}&
\colhead{Scatter}&
\colhead{Sample}&
\colhead{N}
}
 
\startdata
$(5.42 \pm   0.60) +  (-0.153 \pm    0.037)M_B $  & 0.248& KISS \ion{H}{2} Galaxies &        23\\ 
$(5.26 \pm   0.72) +  (-0.168 \pm    0.046)M_B $  & 0.268& Literature \ion{H}{2} Galaxies &  31\\ 
$(5.37 \pm   0.46) +  (-0.159 \pm    0.029)M_B $  & 0.260& All \ion{H}{2} Galaxies    &      54\\   
$(5.59 \pm   0.54) +  (-0.153 \pm    0.025)M_B $  & 0.175& H. Lee et al. (2003) dIrr Galaxies &      22\\   
\enddata 

\end{deluxetable}

\clearpage

\clearpage

\begin{figure}
\epsscale{1.5}
\vspace*{-0.4in}
\hspace*{-0.7in}
\plotone{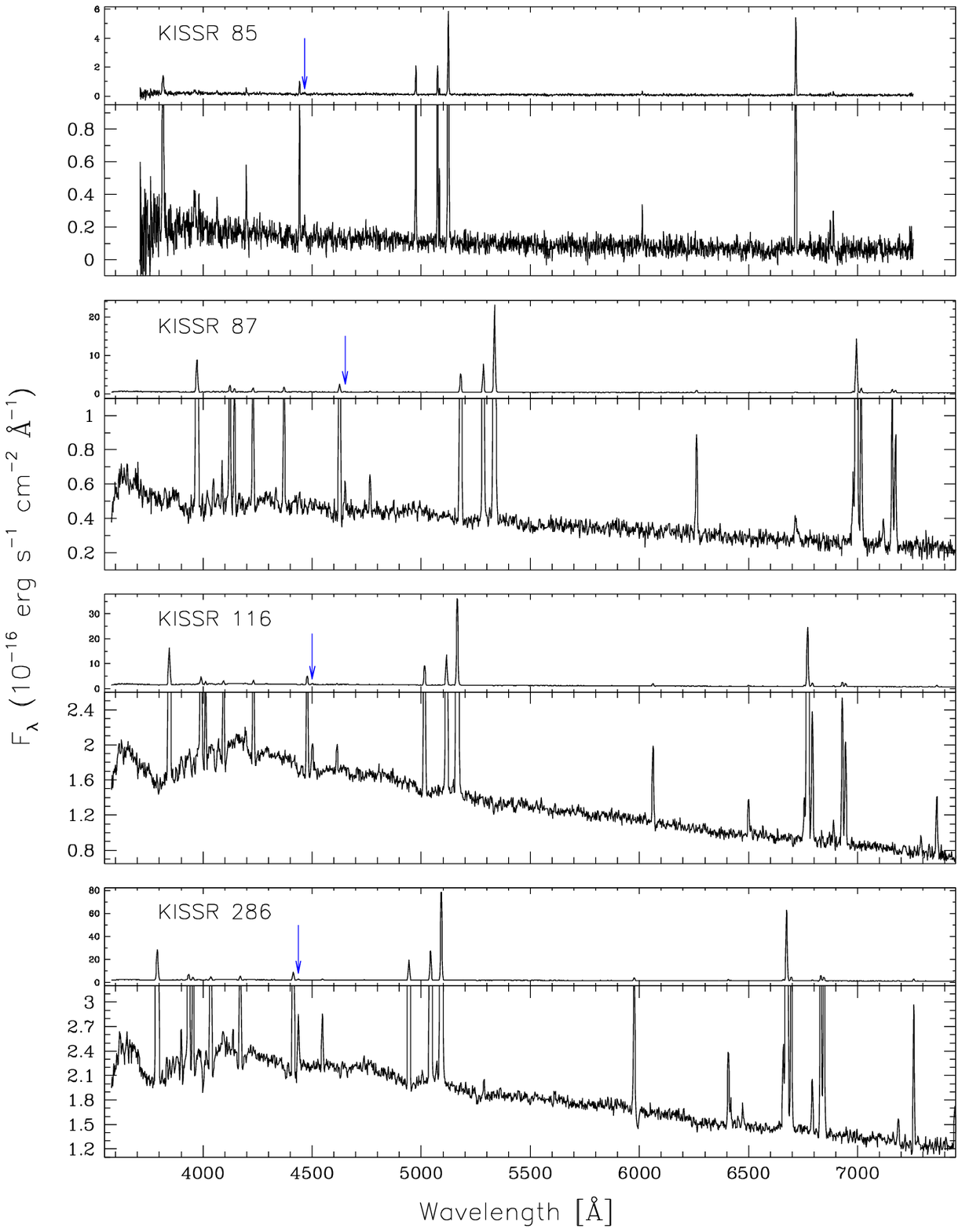}
\end{figure}

\begin{figure}
\epsscale{1.5}
\vspace*{-0.4in}
\hspace*{-0.7in}
\plotone{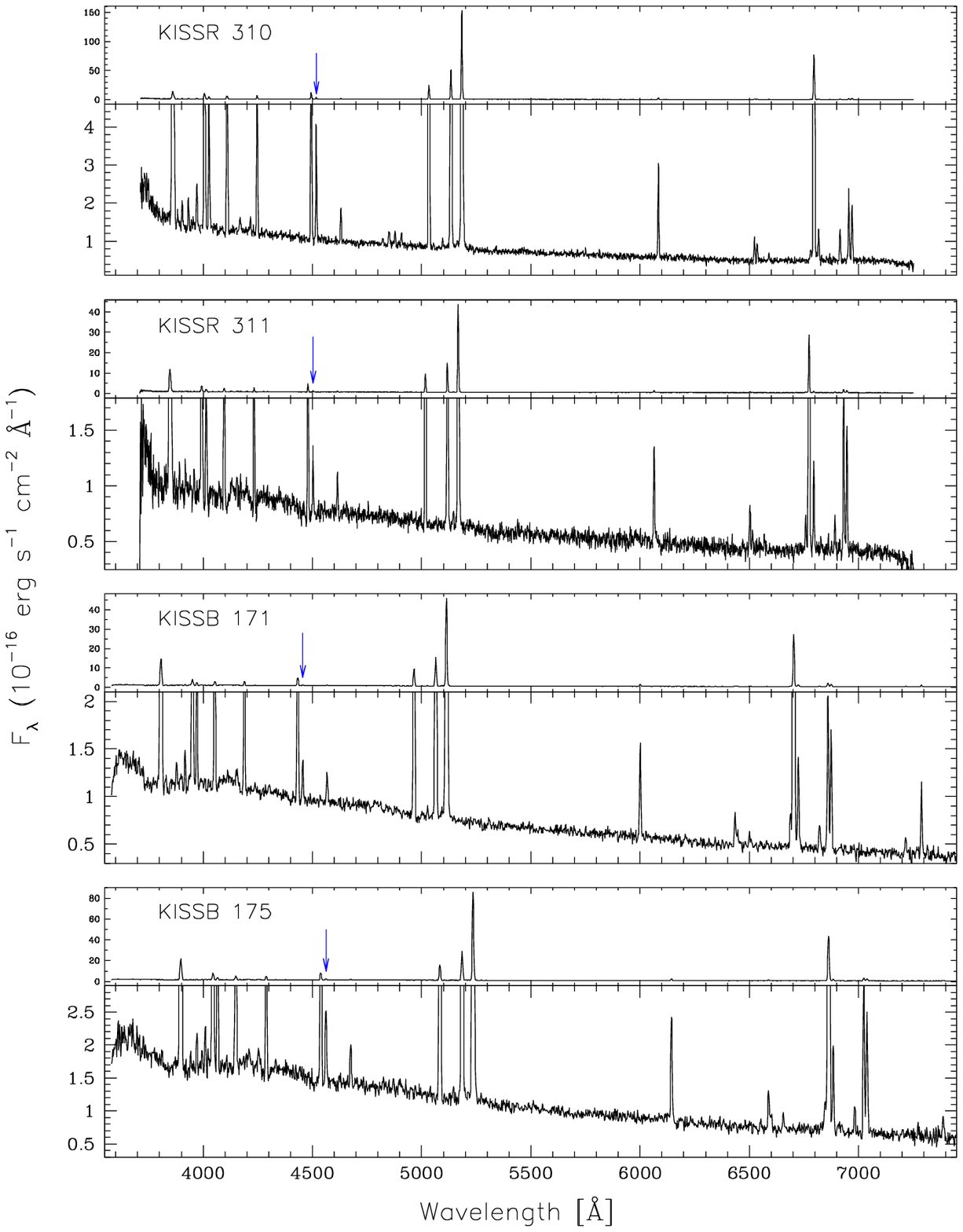}
\end{figure}

\begin{figure}
\epsscale{1.5}
\vspace*{-0.4in}
\hspace*{-0.7in}
\plotone{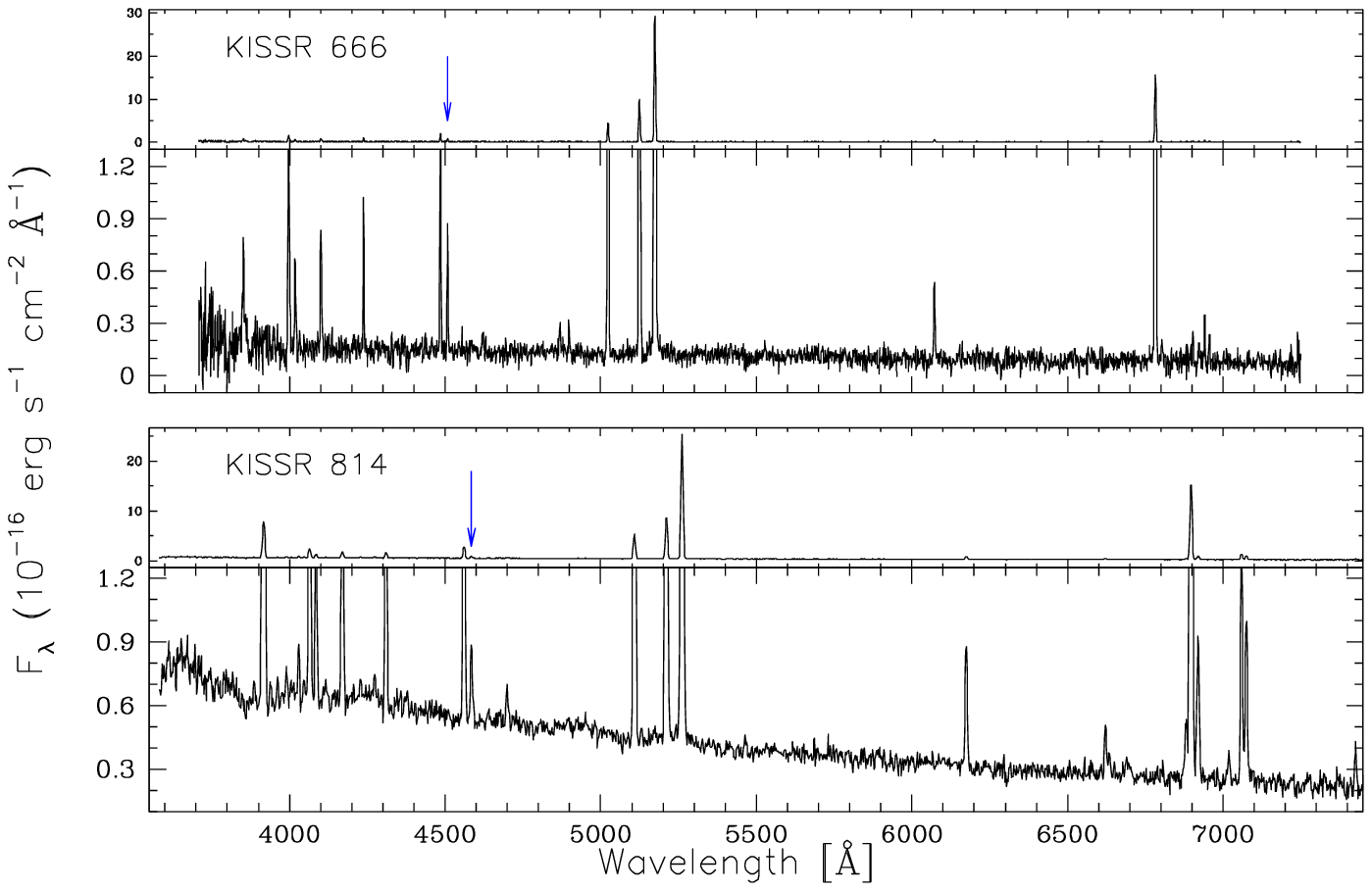}
\end{figure}

\clearpage

\clearpage

\clearpage

\end{document}